\documentclass[twocolumn,prb,showpacs]{revtex4-1}
\usepackage{graphicx}
\usepackage{epstopdf} 
\usepackage{dcolumn}
\usepackage{amsmath}
\usepackage{multirow} 
\usepackage[active]{srcltx}

\begin{document}
\newcommand{\K}{{\vec k}}
\title{Chemical trends of substitutional transition metal dopants in diamond: an \textit{ab initio} study}

\author{T. Chanier, C. Pryor, and M. E. Flatt\'e}
\affiliation{Optical Science and Technology Center and Department of Physics and Astronomy, University of Iowa, Iowa City, Iowa 52242, USA}


\begin{abstract}
The electronic and magnetic properties of neutral substitutional transition-metal  dopants in diamond are calculated within density functional theory using the generalized gradient approximation to the exchange-correlation potential. Ti and Fe are nonmagnetic, whereas the ground state of V, Cr and  Mn are magnetic with a spin entirely localized on the magnetic ion.  For Co, Ni, and Cu, the ground state is magnetic with the spin distributed over the transition-metal ion and the nearest-neighbor carbon atoms; furthermore a bound state is found in the gap that originates from the hybridization of the $3d$-derived level of the dopant and the $2p$-derived dangling bonds of the nearest-neighbor carbons. A $p$--$d$ hybridization model is developed in order to describe the origin of the  magnetic interaction. This model predicts high-spin to low-spin transitions for Ni and Cu under compressive strain.
\end{abstract}


\maketitle 

\section{Introduction}

Diamond has attracted great interest recently for quantum information science because of its wide band gap and the existence of more than 500 optically-addressible centers\cite{Zaitsev2001}, many with long spin coherence times. An extensively-studied color center is the NV center, consisting of a carbon vacancy with a nearest-neighbor (NN) substitutional nitrogen atom. In its negatively-charged state, the NV$^-$ center carries a spin $S=1$, which is split  by the $C_{3v}$ symmetry environment of the spin center\cite{Lenef1996}; its spin state has been proposed as a qubit for quantum computation. Spin initialization and readout has been demonstrated by optical pumping and absorption measurement\cite{Davies1976} at the zero phonon line (ZPL) at 1.945 eV. Spin manipulation has been achieved by microwave manipulation\cite{Jelezko2004b,Hanson2006,Hanson2007}, and also through the interaction of the NV$^{-}$ center spin with a nitrogen nuclear spin \cite{Jelezko2004b} or another nearby substitutional nitrogen electron spin \cite{Hanson2006c}. However, the NV$^-$ center has a large phonon sideband at room temperature that may limit its use for quantum information processing \cite{Jelezko2002}.  

Other applications, like quantum optics and quantum cryptography, require stable, bright, and narrow linewidth single photon sources (SPS). Potential optical centers for SPSs in diamond are the NV$^-$ center, the silicon-vacancy (SiV) center \cite{Su2009,Feng1993}, Ni-related centers \cite{Isoya1990,Isoya1990b,Su2009,Rabeau2005,Orwa2010} and Cr-related centers \cite{Aharonovich2010,Aharonovich2010b}. The large phonon sideband of the NV$^-$ center spreads the emission over 100 nm resulting in a Debye-Waller factor of less than 5\%, which is not ideal for quantum key distribution \cite{Jelezko2002}. The SiV center, formed by a substitutional Si next to a vacancy,  has been proposed as a stable SPS with a ZPL at 738 nm, but its fast nonradiative decay degrades its fluorescence\cite{Feng1993} and ultimately its reliability. 

Transition-metal (TM) dopants have also been studied in diamond, however initial studies focused on TM impurities  in natural diamond samples, or in some synthetic diamond due to the use of the TM as a solvent catalyst during the high-pressure-high-temperature (HPHT) growth process \cite{Iizuka1996,Kiflawi2002,Liu2007,Lin2011}.
Among the Ni-related centers, the NE8 center, formed by a substitutional Ni with four NN nitrogen atoms, is responsible for a ZPL at 793 nm with a narrow phonon sideband, but its low reproducibility impedes its use for applications. Among Cr-related centers, a ZPL at 749 nm \cite{Aharonovich2010} has been attributed to an interstitial chromium and two other ZPL at 744 and 764 nm have been detected but the source has not yet been identified\cite{Aharonovich2010b}.  TM dopants can also be introduced by ion implantation  \cite{Orwa2010,Aharonovich2010} or during chemical vapor deposition (CVD)  \cite{Rabeau2005,Aharonovich2010b}, which offer the possibility of introducing the dopants substitutionally and without additional atoms or as part of a defect complex. Co- and Ni-related centers, with different nitrogen, boron and vacancy complexes, have been studied in  various charge states by GGA calculations\cite{Larico2008,Larico2009}, finding neutral substitutional Co to be stable in pure diamond for a Fermi energy $E_F$ between 3.0 and 3.6 eV, and neutral substitutional Ni to be stable for $E_F$ between 2.6 and 3.0 eV.

Transition-metal spin centers in diamond offer several potential advantages over other spin centers, including (1) the availability of both strongly-hybridizing ($t_2$-symmetry) and weakly-hybridizing ($e$-symmetry) states associated with the $d$ levels\cite{Hjalmarson1980,Vogl1985}, (2) the high symmetry of the substitutional dopant (tetrahedral group, which means the crystal field does not split angular-momentum $1$ triplets\cite{Yu3ed}), and (3) the availability of larger spin-orbit interactions through the $d$ electrons, which may permit faster manipulation of a spin through an electric field (\textit{e.g.} Ref.~\onlinecite{Tang2006}). Challenges for these spin centers include the large strain introduced when they are placed substitutionally (although formation energies have previously been found to be moderate), and the potential lower spin coherence time due to the larger spin-orbit interaction. 

To clarify these phenomena we present a first-principles study of substitutional transition-metal (TM$_s$) dopants in diamond. Whereas TM$_s$ dopants may exist in several different charge states in diamond, we concentrate our study on the neutral substitutional TM dopant (TM$^0_s$) and the chemical trends within the $3d$ row of the periodic table. For each TM$^0_s$ dopant we determine the ground-state configuration, including the charge and spin state. We find for Ti through Cu a single $p$--$d$ hybridization model explains the electronic and magnetic properties of the defect. Sc and Zn, which occur at the ends of the row, are not described by this model. In Section~\ref{results}, after a description of the method,we present the results for  $3d$ transition-metal dopants,  in Section~\ref{highspin-lowspin} we describe  changes under strain of the ground-state spin for Ni and Cu and contrast with Cr (which is largely unchanged by strain). In Section~\ref{conclude} we comment on the implications of these results and the strain-induced spin changes for quantum information processing.

\section{Dopant Configuration Calculations  and $p-d$ Hybridization Models}\label{results}

\subsection{Method}

The calculations were performed with the scalar relativistic version of the full potential local orbital FPLO9.00-33 code \cite{Koepernik1999} using the (spin-polarized) generalized gradient approximation ((S)GGA) with the parametrization of Perdew, Burke
and Ernzerhof \cite{Perdew1996}. The convergence of the results with respect to ${\bf k}$-space integrals was carefully checked. We found that a 8x8x8 {\bf k}-point grid were sufficient to obtain convergence of the total energy difference between spin-polarized (SGGA) and spin unpolarized (GGA) calculations. 

We used a 64-atom TMC$_{63}$ (TM = Sc, Ti, V, Cr, Mn, Fe, Co, Ni, Cu, Zn) supercell  corresponding to a $2\times 2 \times 2$ multiple of the diamond conventional cell with  a TM substituted for one carbon. 
We fixed the supercell to correspond to the experimental lattice constant\cite{Riley1944} of diamond $a_0=3.5668$ \AA, and all the 
atomic positions within the supercell were allowed to relax with a precision of 1 meV/\AA.  We considered here only 
 $T_d$-symmetric atomic relaxation. As a check, we did a fully-relaxed-atoms calculation without any symmetry constraints ($C_1$)for the NiC$_{63}$ supercell. The $C_1$ relaxed calculation is about 2 meV higher in energy than the $T_d$ relaxed calculation, which rules out such a symmetry-breaking distortion. 

\subsection{Transition-metal dopant ground-state spin configurations}

The SGGA results are summarized on Table \ref{TMchemtrend}. The most stable magnetic solution is obtained for Cr$^0_s$, with a magnetic energy $E_M=-1006$ meV, corresponding to an $S=1$ ground state  $E_M$  lower in energy than the non-magnetic GGA solution. Less stable magnetic solutions ($E_M\sim -250$~meV) are obtained for the $3d$ neighbors V$^0_s$  and Mn$^0_s$, which we shall see below become magnetic in a similar fashion to Cr$^0_s$. A different origin is found for magnetism in  Co$^0_s$ ,Ni$^0_s$ and Cu$^0_s$, which have much smaller magnetic energies $\sim-100$ meV. Sc$^0_s$ and Zn$^0_s$ are included to complete the $3d$ row, however the magnetism of these ions differs from that of all the other $3d$ ions.

\begin{table}[h!]
\begin{tabular}{|c|c|c|c|c|c|c|c|c|c|c|c|c|c|}\hline
TM & $R$[$\%$] & $E_\mathrm{M}$[meV] & $M_{T}[\mu_\mathrm{B}$] & $M_\mathrm{TM}$[$\mu_\mathrm{B}$] & $M_\mathrm{C}$[$\mu_\mathrm{B}$] & $N_\mathrm{TM}$ & $N_\mathrm{C}$\\\hline
Sc &    23.2 & -49 &  1.0 &  0.2 &  0.2 &  19.3 & 6.4 \\\hline
Ti &    19.0 & 0 &  0.0 &  0.0 &  0.0 &  20.1 & 6.5 \\\hline
V &     16.7 &   -249 & 1.0 & 1.0 &  0.0 &  21.2 & 6.5  \\\hline
Cr &    14.8 &  -1006 & 2.0 & 2.0 & 0.0 &  22.3 & 6.4  \\\hline
Mn &    13.3 &   -246 & 1.0 & 1.0 &  0.0 &  23.5 & 6.4 \\\hline
Fe &    12.2 & 0 &  0.0 &  0.0 &  0.0 &  24.6 & 6.4 \\\hline
Co &    12.9 &  -42 & 1.0 & 0.6 &  0.1 &  25.9 & 6.3 \\\hline
Ni &    14.2 &   -71 & 2.0 & 0.9 &  0.2 &  26.9 & 6.3 \\\hline
Cu &    15.7 &  -127 & 3.0 & 0.8 &  0.4 &  28.1 & 6.2 \\\hline
Zn &    16.6 & -100 &  2.0 &  0.2 &  0.3 &  29.0 & 6.3 \\\hline
\end{tabular}
\caption{TMC$_{63}$ SGGA  Results : $R$ is the NN relaxation, $E_M=E_\mathrm{SGGA}-E_\mathrm{GGA}$ the magnetic energy, $M_T$ the total magnetization, $M_\mathrm{TM}$ ($M_\mathrm{C}$) the on-site magnetization on the TM ion (NN C atom), $N_\mathrm{TM}$ ($N_\mathrm{C}$) the calculated number of electrons on the TM ion  (NN C atom).}\label{TMchemtrend} 
\end{table}

\begin{figure}[h!]
\begin{center}
\includegraphics[width =85mm]{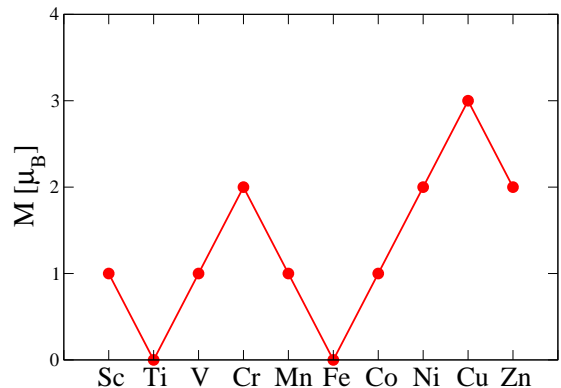}\\[-0.7cm]
\caption{TMC$_{63}$ SGGA total magnetic moment.}\label{MvsTM} 
\end{center}
\end{figure}

Figure \ref{MvsTM} shows the calculated total magnetization of TMC$_{63}$ supercells. Ti and Fe are non-magnetic. We obtain for Cr and Ni a total magnetization $M_T=2\ \mu_\mathrm{B}$, corresponding to a total spin $S=1$, although we shall see later that these two spin centers differ greatly due to the importance of different $d$ orbitals in creating the ground-state spin. Zn, also with $S=1$, differs from both as it has a closed $d$ shell. For Sc, V, Mn and Co, we obtain a total magnetization $M_T=1\ \mu_\mathrm{B}$, corresponding to a total spin $S=\frac{1}{2}$. The highest total magnetization is obtained for Cu with $M_T=3\ \mu_\mathrm{B}$ corresponding to a total spin $S=\frac{3}{2}$.

 For V$^0_s$, Cr$^0_s$ and Mn$^0_s$, the total magnetization of $M_T=$~1.0, 2.0, and 1.0~$\mu_\mathrm{B}$ is entirely localized on the TM ion, whereas for Co$^0_s$, Ni$^0_s$ and Cu$^0_s$, the total magnetization of $M_T=$~1.0, 2.0, and 3.0~$\mu_\mathrm{B}$ is distributed between the TM ion and the NN carbon atoms with a TM ion magnetization of $M_\mathrm{TM}=$~0.6, 0.9 and 0.8~$\mu_\mathrm{B}$ and a  NN carbon atom magnetization of $M_\mathrm{C}=$~0.1, 0.2 and 0.4~$\mu_\mathrm{B}$, corresponding to a total magnetic moment 0.4, 0.8 and 1.6~$\mu_\mathrm{B}$ distributed over the NN C. The number of electrons located on the TM ion   compared to its atomic number, and the number of electrons on the NN carbon atoms allow us to identify the configuration of the TM ion and the induced charge on the NN C's. 
 
 From Ti to Fe, the TM occupation number is consistent with a TM$^{2+}$ configuration $4s^03d^n$ ($n$=2,3,4,5,6) and the TM$^{2+}$ ion induces two extra electrons distributed over the 4 NN carbon dangling bonds with a local charge on each NN carbon atoms $Q_\mathrm{C}\sim0.5$. For Co, Ni, Cu, the TM occupation number is consistent with a TM$^{+}$ configuration $4s^03d^{n+1}$ ($n$ = 7, 8, 9) and the TM$^+$ ion induces one extra electron distributed over the 4 NN carbon dangling bonds with a local charge on each NN carbon atoms $Q_\mathrm{C}\sim0.2$. Sc and Zn differ greatly from all the other $3d$ dopants, and hence will be discussed only at the end of Section~\ref{results}.

\begin{figure}[h!]
\begin{center}
\includegraphics[width =87mm]{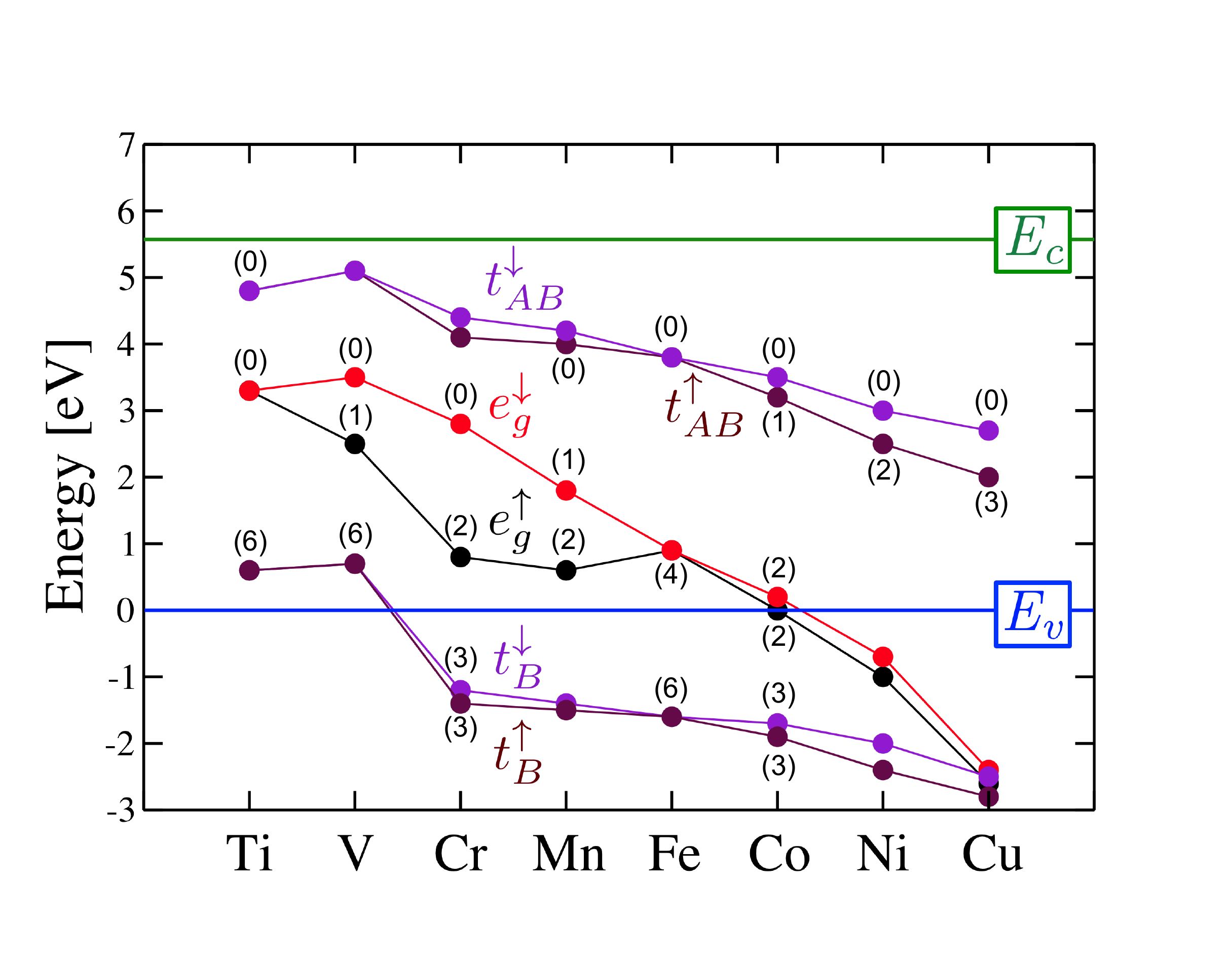}\\
\caption{Single-particle TM $3d$-derived SGGA Kohn-Sham (KS) energy levels of TMC63 calculation. TM $3d$ levels are split by the tetrahedral crystal field into $e^{\sigma}$ and $t_{2}^{\sigma}$ levels. The bonding and antibonding levels ($t_B^\sigma$ and $t_{AB}^\sigma$) are the results of the hybridization between the TM $t_{2}$ and NN carbon $2p$-derived $t_2$ levels. $E_v$ (energy reference) and $E_c$ are the SGGA valence band maximum and (direct) conduction band minimum of pure diamond. In parenthesis are indicated the occupation number of the KS energy levels.}\label{KSvsTM}
\end{center}
\end{figure}

The TM $3d$-derived Kohn-Sham states within the series are presented in Figure \ref{KSvsTM}. The GGA direct and indirect band gap for pure diamond are respectively $E_{GGA}^{D}=E_c-E_v=5.6$ and $E_{GGA}^{Ind}=4.3$ eV, both of which are smaller than the experimental  gap $E_G^{Ind}=5.47$ eV due to the well known gap underestimation of the GGA functional \cite{Filippi1994}. A substitutional TM ion is at the center of a tetrahedron formed by the 4 NN carbon atoms,  implying a crystal field of $T_d$ symmetry. This $T_d$ crystal field will split the single particle $3d$ manifold into a doubly degenerate $e$ level and a  triply degenerate $t_{2}$ level. For tetrahedral bonding the $t_{2}$ levels are expected to be higher in energy than the $e$ levels. The symmetry of the bonds from the four nearest neighbors is compatible with $t_2$ symmetry\cite{Hjalmarson1980}, but not with $e$ symmetry. Thus the $3d$-derived $t_{2}$ level hybridizes with the NN carbon $2p$-derived $t_2$ states to form bonding $t_B$ and antibonding $t_{AB}$ hybrid states. Due to the strong hybridization between the TM ion and the NN carbon atoms, the $3d$-derived hybrid states will form bound states in the gap \cite{Vogl1985,Chanier2009,Dietl2008}. 

For the non-magnetic Ti$^0_s$ impurity, there are 3 bound states in the gap : a totally occupied (6-fold degenerate) $t_B$ level situated at $E_v+0.6$ eV, a totally empty (4-fold degenerate) $e$ level at $E_v+3.3$ eV and a totally empty  (6-fold degenerate) $t_{AB}$ level at $E_v+4.8$ eV. For V$^0_s$, Cr$^0_s$, Mn$^0_s$, the spin up and spin down (3-fold degenerate) $t_B$ levels are totally occupied and nearly spin degenerate. For V, the $t_B$ forms a bound state at $E_v+0.7$ eV and for Cr$^0_s$ to Cu$^0_s$, the $t_B$ state forms a crystal field resonance state inside the valence band which (with increasing atomic number) decreases in energy. For V, Cr and Mn, the occupation difference of the spin up and spin down $e$ bound states is responsible for the total magnetization localized on the TM ion; the Cr $e$ level configuration is $(e^\uparrow)^2(e^\downarrow)^0$, corresponding to a total magnetic moment $M_T=2.0\ \mu_\mathrm{B}$ ($S=1$). The absence of magnetism for Fe is due to a totally occupied spin-degenerate $e$ level (4-fold degenerate). For the next three dopants in the series (Fig.~\ref{MvsTM}), Co, Ni and Cu, the $e$ levels remain totally occupied and enter the valence band. For these dopants the spin up antibonding (3-fold degenerate) $t_{AB}^\uparrow$ state starts to become occupied, with 1, 2 or 3 electrons respectively, while the spin down antibonding $t_{AB}^\downarrow$ level remains totally empty. For Co$^0_s$, Ni$^0_s$ and Cu$^0_s$, the occupation difference of the spin up and spin down $t_{AB}$ bound states is responsible for the total magnetization that is distributed over the TM ion and the NN carbon atoms through the hybrid $t_{AB}$ level; for example, the Ni $t_{AB}$ level configuration is $(t_{AB}^\uparrow)^2(t_{AB}^\downarrow)^0$, corresponding to a total magnetic moment $M_T=2.0\ \mu_\mathrm{B}$ ($S=1$) distributed over the Ni ion and the NN carbon atoms. 

A $p$--$d$ hybridization model\cite{Hjalmarson1980,Vogl1985} helps explain the origin of the magnetic interaction and differentiate the behavior of those dopants to the left of Fe from those to the right. In pure diamond, each carbon shares 4 electrons to form covalent bond. If one carbon atom is removed to form a vacancy, 4 electrons will occupy the dangling bonds with the  neutral configuration $2s^22p^2$. When a TM ion is inserted at the vacancy site to form a substitutional dopant, it will induce an extra charge on the dangling bond distributed over the 4 NN carbons.  For TM=V,Cr and Mn, the magnetization is driven by the $e$ bound states with the magnetization entirely localized on the TM ion in the TM$^{2+}$ configuration $4s^03d^n$ and two extra electrons shared by the non-magnetic NN carbon atoms. These two extra electrons will occupy the dangling bonds and form a defect level of configuration $2s^22p^4$. The TM $3d$-derived $t_2$ levels hybridize with the $2p$-derived $t_2$ states occupied by 4 electrons. For TM=Co,Ni and Cu, the magnetization is driven by the antibonding $t_{AB}$ bound states. The TM is in the TM$^{+}$ configuration $4s^03d^{n+1}$ which induces one extra electron distributed over the NN carbon atoms. This extra electron will occupy the dangling bonds and form a defect level of configuration $2s^22p^3$ which induces a non-zero magnetic moment on the NN carbon atoms. The TM $3d$-derived $t_2$ levels will hybridize with these $2p$-derived $t_2$ states occupied by 3 electrons. 

In the next subsections, we will present the electronic structure of the different TM$^0_s$ dopants sorted by their spin values. Dopants with the same spin values can have dramatically-different origins for the magnetism, depending on whether the magnetism is driven by the $e$ states or the $t_{AB}$ states. This difference will also lead to a very different response of the ground-state spin to strain.

\subsection{The $S=0$ centers : Ti$^0_s$ and Fe$^0_s$}

\begin{figure}[h!]
\begin{tabular}{|c|}
\hline
\includegraphics[width = 85mm]{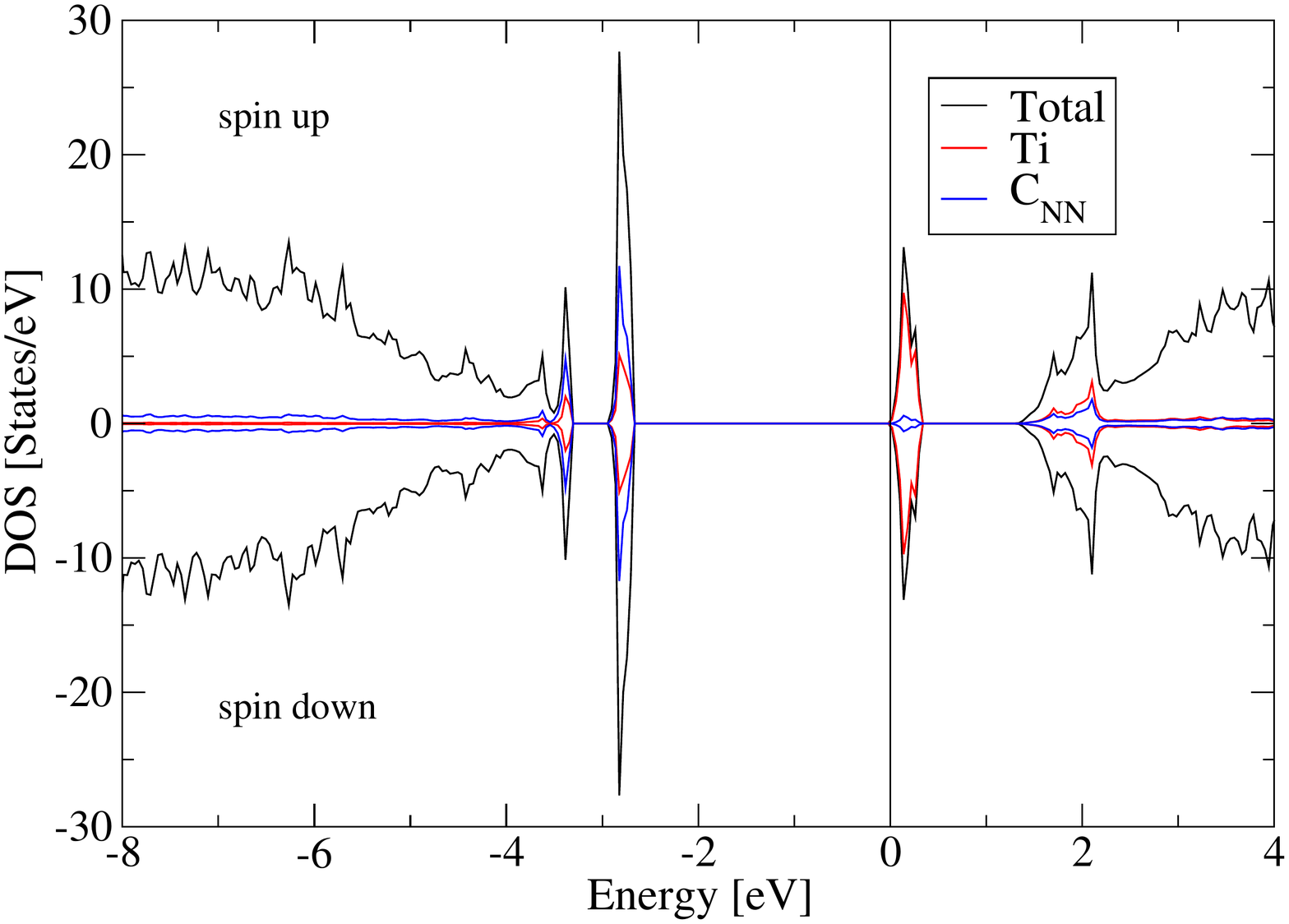}\\\hline  
 \\
\includegraphics[width = 80mm]{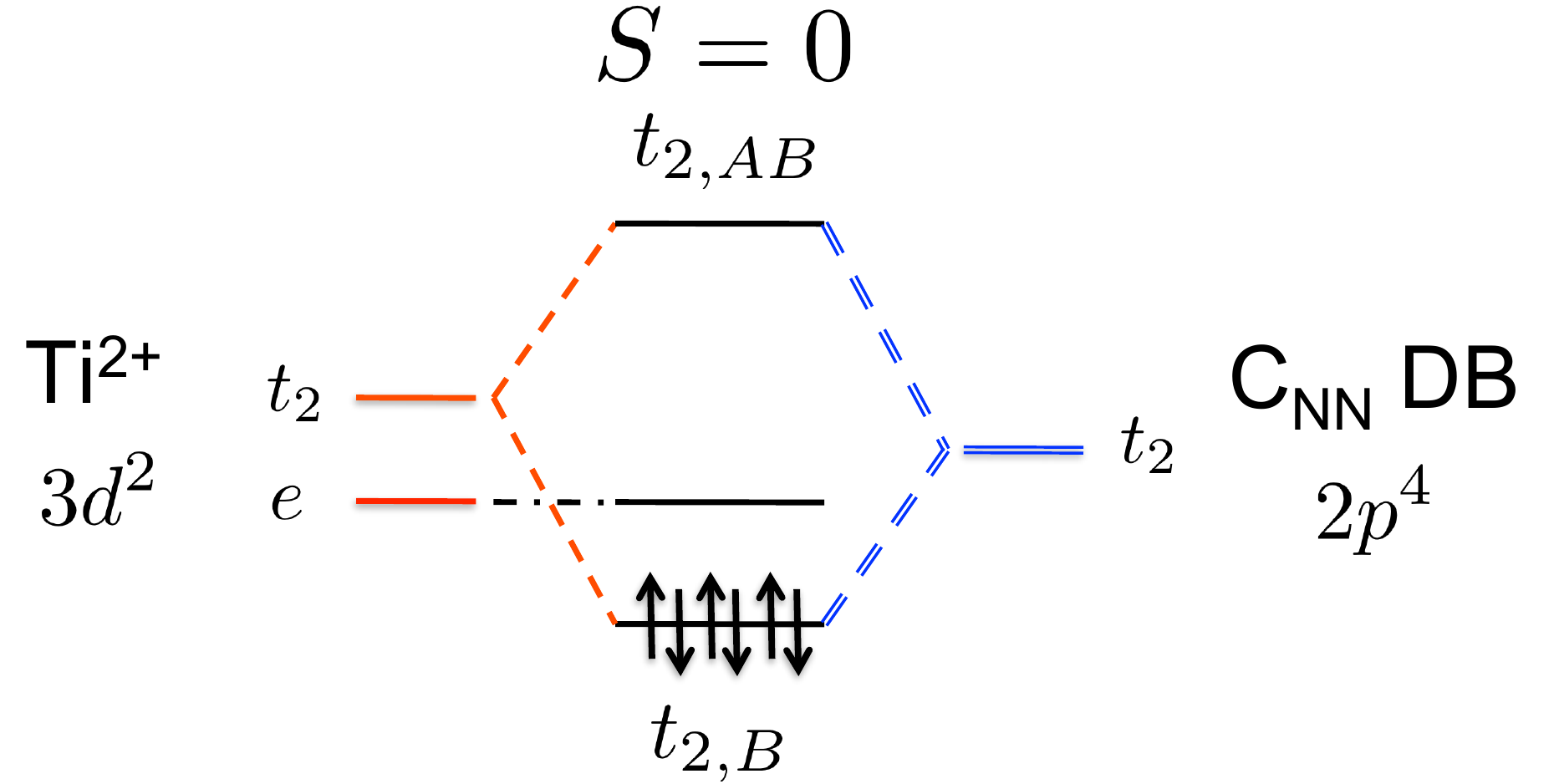}\\  
 \\\hline
 \end{tabular}
\caption{(color online)  Top : Total (black solid line), Ti (red dashed line) and NN C (blue, dot-dashed line) partial DOS. For these and subsequent such plots the upper portion of the plot is for the majority spin and the lower portion is for the minority spin; here the partial DOS's are the same as the ion is nonmagnetic. The bonding state $t_B$ is delocalized across the  Ti $3d$ and NN C $2p$ states. Bottom : Schematic representation of the $p$--$d$ hybridization model for Ti$^0_s$. 
}\label{Ti} 
\end{figure}

Fig. \ref{Ti} shows the total, Ti and NN carbon partial density of states (DOS) of the  TiC$_{63}$ supercell. The Ti $3d$ levels are split by the $T_d$ crystal field  into $e$ and $t_2$ levels. Ti $t_2$ levels hybridize with the NN carbon $2p$ levels to form bonding $t_B$ and antibonding $t_{AB}$ hybrid levels. The spin-degenerate bonding $t_B$ levels are totally occupied and form a bound state at 0.6 eV above the valence band maximum (VBM). The spin-degenerate $e$ levels are totally empty and form a bound state at 3.3 eV above the VBM. The  spin-degenerate antibonding $t_{AB}$ levels  are totally empty and form a bound state situated at 4.7 eV above the VBM. The number of electrons located on the Ti ion is 20.1 which corresponds to a Ti$^{2+}$ configuration $4s^03d^2$ and spin $S_1=0$. Ti$^{2+}$ induces two extra electrons distributed over NN carbon dangling bonds of configuration $2s^22p^4$ and spin $S_2=0$. The Ti $(t_2)^2$ level hybridizes with the  NN carbon $2p^4$ $t_2$ defect level. The spin-degenerate bonding $t_B$ states are totally occupied and the $e$ levels are totally empty, which explains the absence of magnetism. Fig. \ref{Ti} explains the observed DOS by a $p$--$d$ hybridization model. The two Ti $3d$ electrons occupy the spin-degenerate Ti $t_2$ level (violating Hund's rule) to form a lower energy closed-shell bonding $t_B$ level with the four $t_2$ electrons contributed by the nearest-neighbor carbons.

\begin{figure}[h!]
\begin{tabular}{|c|}
\hline
\includegraphics[width = 85mm]{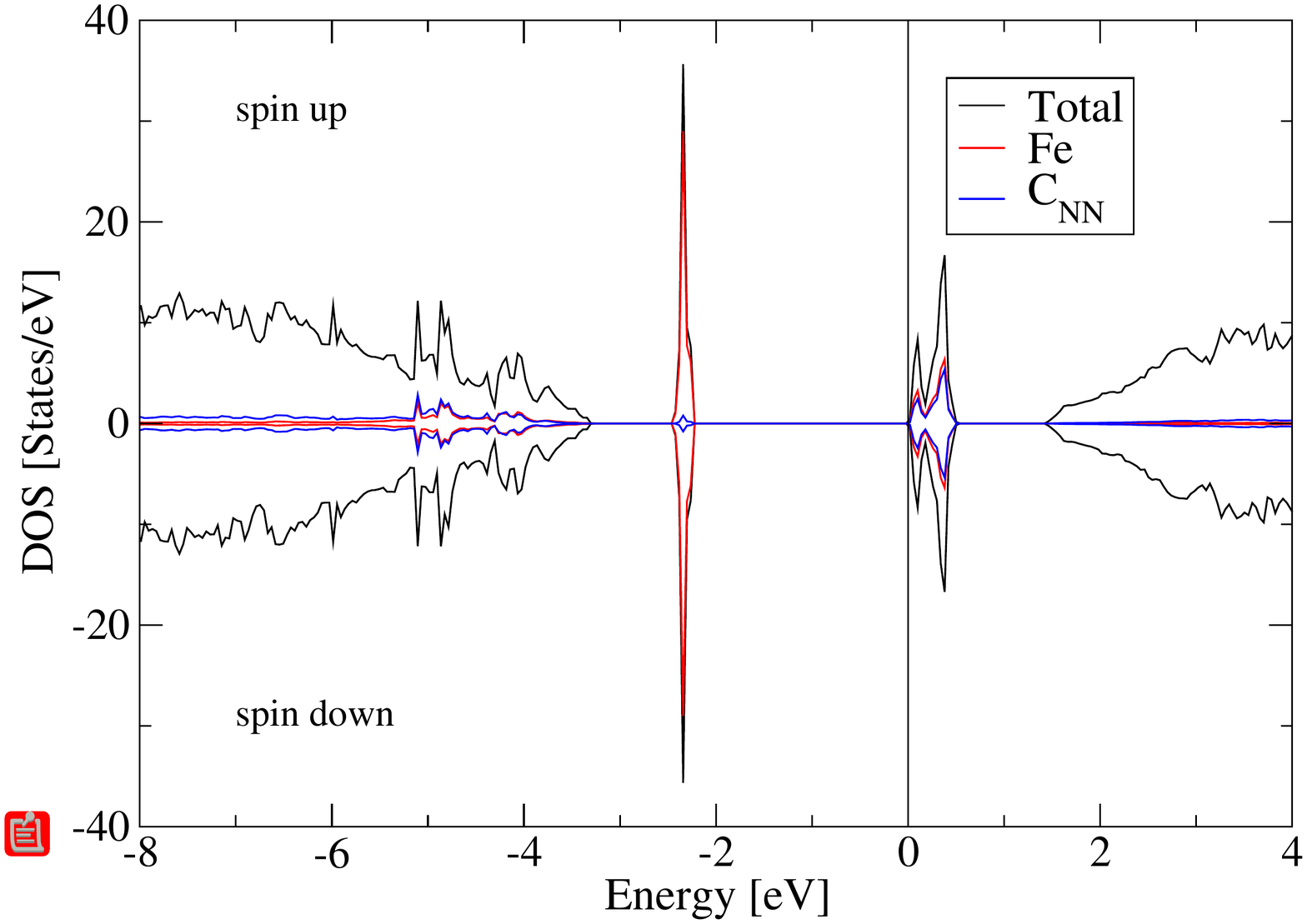}\\\hline  
 \\
\includegraphics[width = 80mm]{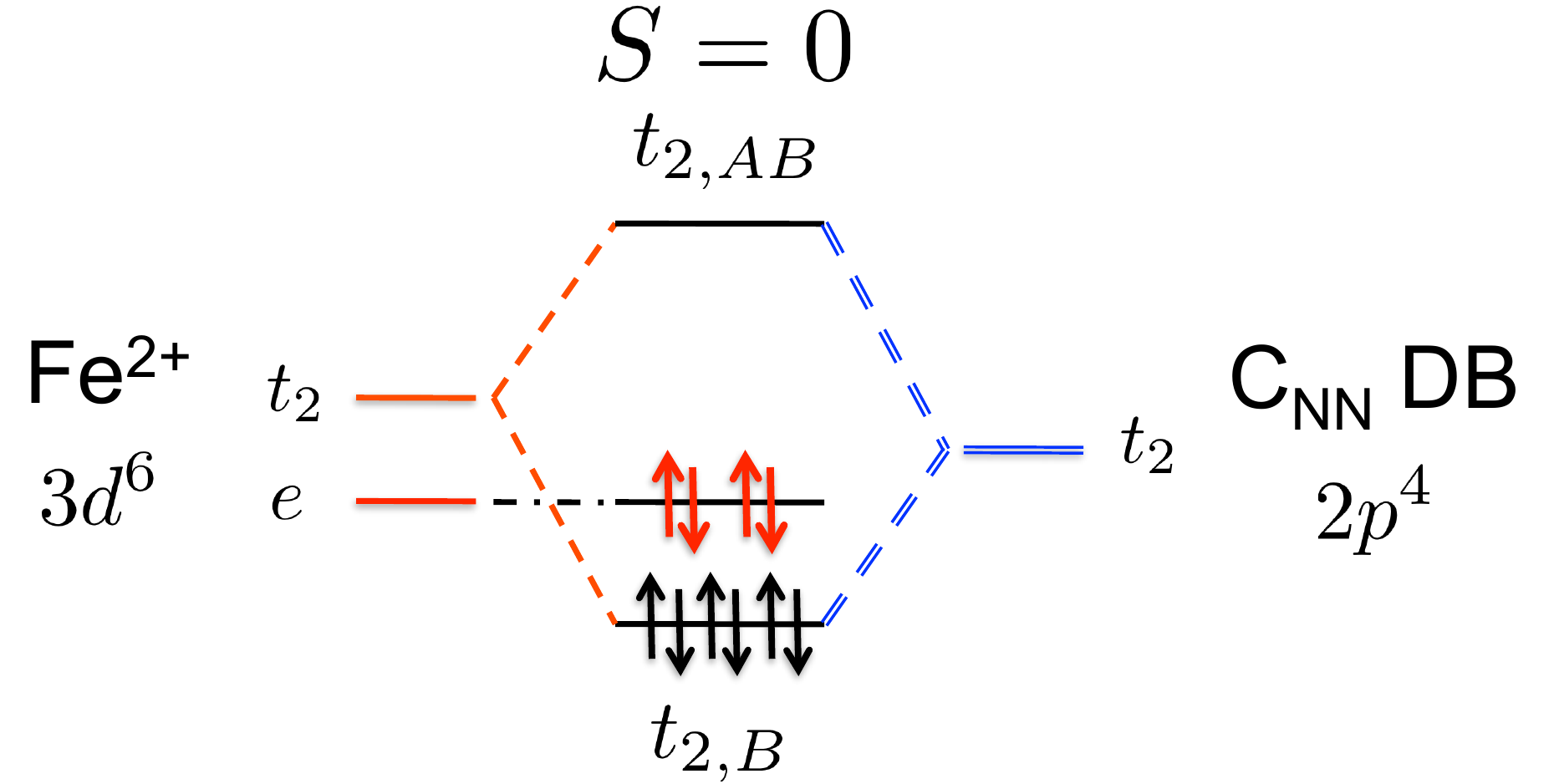}\\  
 \\\hline
 
 \end{tabular}
\caption{(color online) Top : Total, Fe and NN C partial DOS. Bottom : Schematic representation of the $p$--$d$ hybridization model for Fe$^0_s$. The $e$ states are indicated in red to emphasize their localized character; $t_2$ states are indicated in black and are delocalized between the Fe and NN C. Symbols are the same as in Fig.~\ref{Ti}.}\label{Fe} 
\end{figure}

Fig. \ref{Fe} shows the total, Fe and NN carbon partial DOS of the  FeC$_{63}$ supercell. The Fe $3d$ levels are split by the $T_d$ crystal field   into $e$ and $t_2$ levels. Fe $t_2$ levels hybridize with the NN carbon $2p$ levels to form bonding $t_B$ and antibonding $t_{AB}$ hybrid levels.  The spin-degenerate bonding $t_B$ levels are totally occupied and form a crystal field resonance state at -1.7 eV below the VBM. The spin-degenerate $e$ levels are totally occupied and form a bound state at 0.9 eV above the VBM. The spin-degenerate antibonding $t_{AB}$ levels are totally empty and form a bound state situated at 3.8 eV above the VBM. The number of electrons located on Fe is 24.6, which corresponds to a  Fe$^{2+}$ configuration of $4s^03d^6$ and spin $S_1=0$. Fe$^{2+}$ induces two extra electrons distributed over the NN carbon dangling bonds, which have a  $2s^22p^4$ configuration and spin $S_2=0$. The Ti $(t_2)^2$ level hybridizes with the  NN carbon $2p^4$ $t_2$ defect level. The spin-degenerate bonding $t_B$ and  $e$ levels are totally occupied explaining the absence of magnetism. Fig. \ref{Fe} explains the observed DOS by a $p$--$d$ hybridization model, showing it is very similar to Ti, but with a full $e$ level instead of the empty $e$ level of Ti.

\subsection{The $S=\frac{1}{2}$ centers : V$^0_s$, Mn$^0_s$ and Co$^0_s$.}

\begin{figure}[h!]
\begin{tabular}{|c|}
\hline
\includegraphics[width = 85mm]{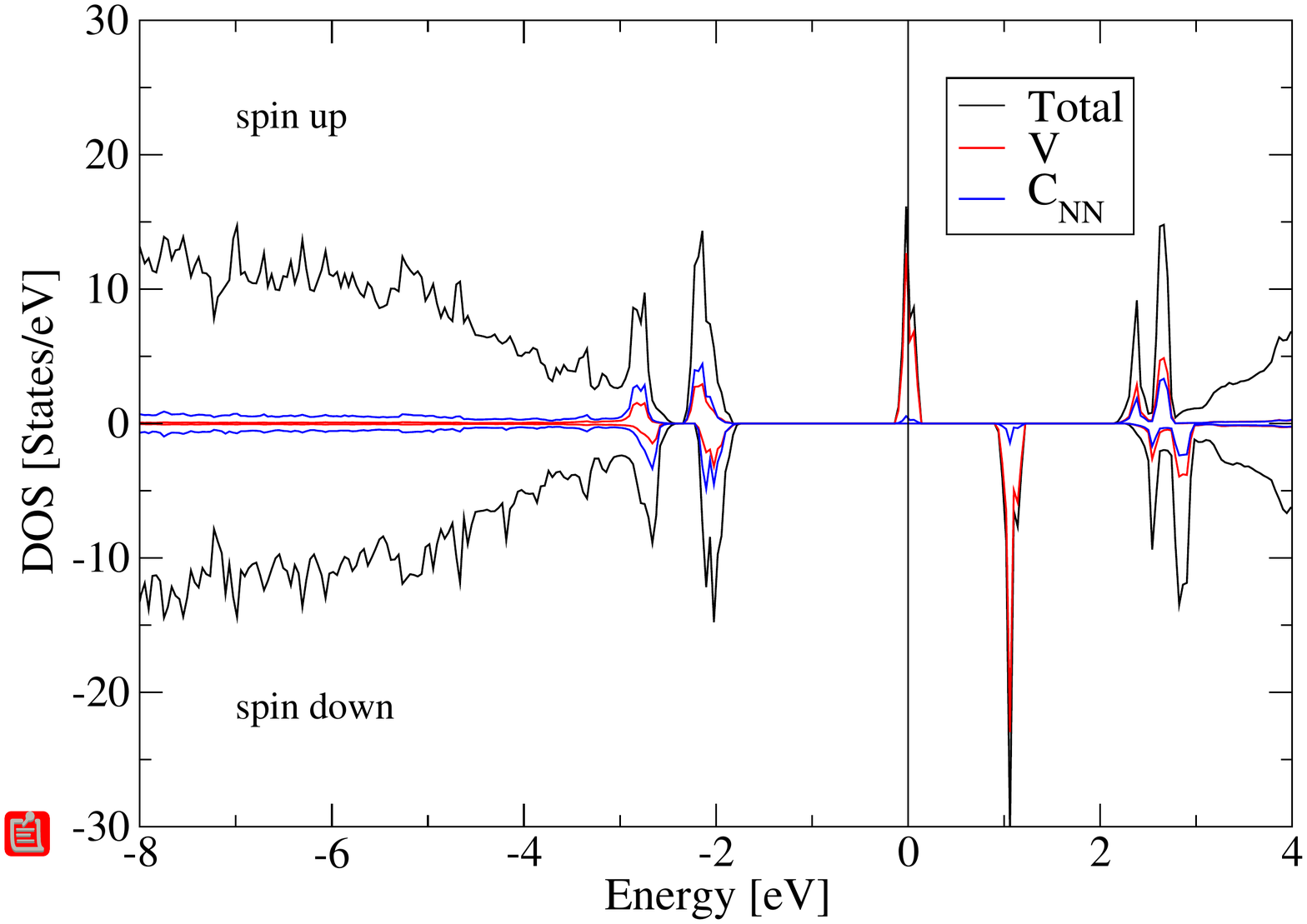}\\\hline  
 \\
\includegraphics[width = 80mm]{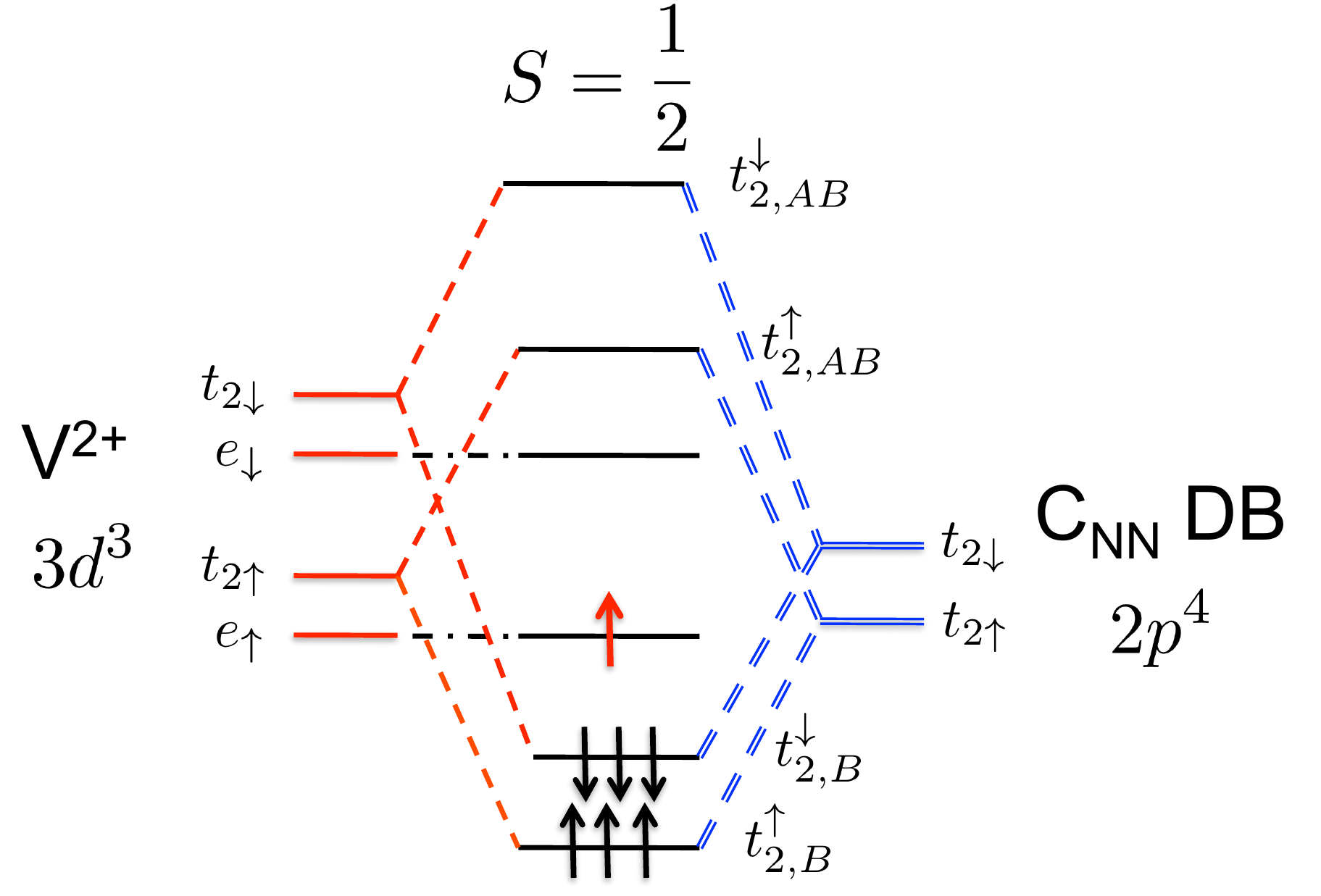}\\  
 \\\hline
 
 \end{tabular}
\caption{(color online) Top : Total, V and NN C partial DOS.  As before the upper portion of the plot is for the majority spin and the lower portion is for the minority spin. Bottom : Schematic representation of the $p$--$d$ hybridization model for V$^0_s$. Symbols are the same as in Fig.~\ref{Ti}.}\label{V} 
\end{figure}

Fig. \ref{V} shows the total, V and NN carbon partial DOS of the VC$_{63}$ supercell. The V $3d$ levels are split by the $T_d$ crystal field  into $e$ and $t_2$ levels. V $t_2$ levels hybridize with the NN carbon $2p$ levels to form bonding $t_B$ and antibonding $t_{AB}$ hybrid levels. The bonding $t_B$ levels are totally occupied and form a nearly spin-degenerate bound state at 0.7 eV above the VBM. The spin up and spin down $e$ levels are localized on the V site, giving rise two bound states situated at 2.5 eV and 3.5 eV above the VBM  and separated by a Hund exchange splitting $J^H=1$ eV. The V $e$ level contribution to the total magnetization is about 60 \% and the bonding $t_B$ states are weakly spin-polarized. The antibonding $t_{AB}$ states are totally empty and form a nearly spin-degenerate bound state at 5.1 eV above the VBM. The number of electrons located on the V ion is 21.2, which corresponds to the V$^{2+}$ configuration $4s^03d^3$ and spin $S_1=\frac{1}{2}$. V$^{2+}$ induces two extra electrons distributed over the NN carbon dangling bonds of configuration $2s^22p^4$ and spin $S_2=0$. The V $(t_2)^2$ level hybridizes with the  NN carbon $2p^4$ $t_2$ defect level. The bonding $t_B$ states are fully occupied (as for Ti and Fe), but in the case of V the $e$ levels are partially occupied in the configuration $(e^\uparrow)^1(e^\downarrow)^0$ corresponding to a total magnetic moment $M_T=1.0\ \mu_\mathrm{B}$ entirely localized on V. Fig. \ref{V} explains the observed DOS with a $p$--$d$ hybridization model.

\begin{figure}[h!]
\begin{tabular}{|c|}
\hline
\includegraphics[width = 85mm]{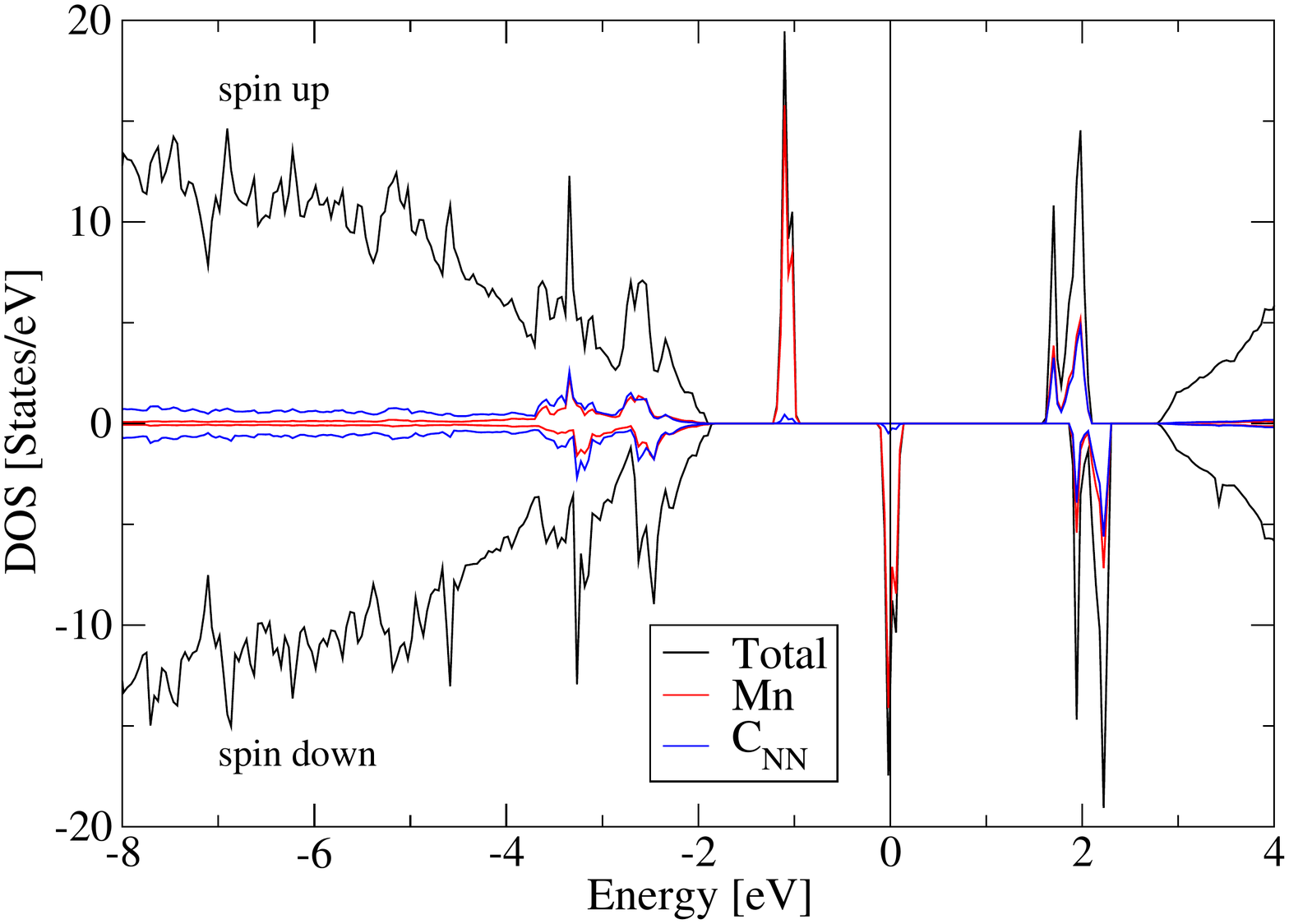}\\\hline  
 \\
\includegraphics[width = 80mm]{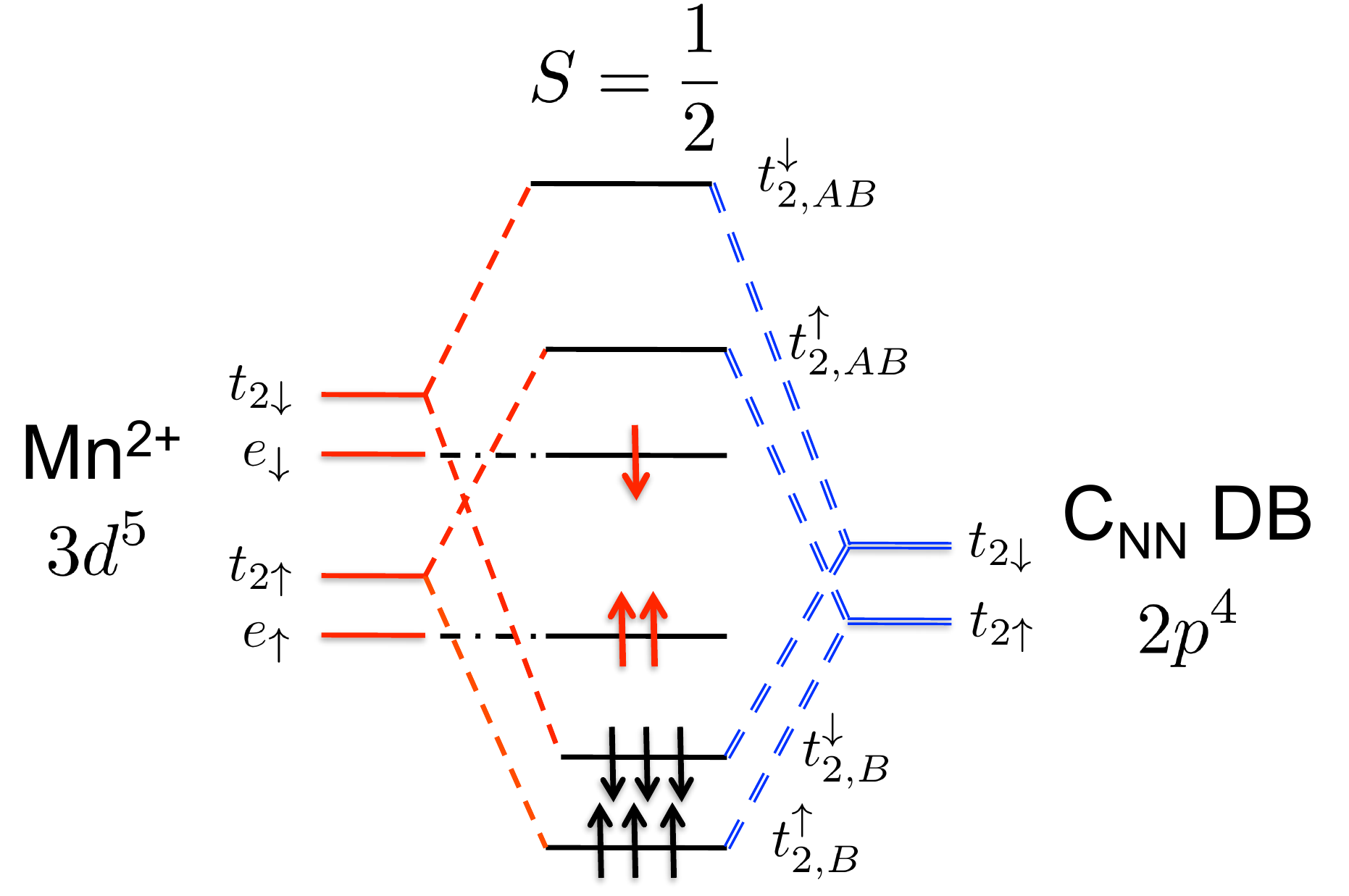}\\  
 \\\hline
 
 \end{tabular}
\caption{(color online) Top : Total, Mn and NN C partial DOS. Bottom : Schematic representation of the $p$--$d$ hybridization model for Mn$^0_s$. Symbols are the same as in Fig.~\ref{Ti}.}\label{Mn} 
\end{figure}

Fig. \ref{Mn} shows the total Mn and NN carbon partial DOS of the MnC$_{63}$ supercell. The Mn $3d$ levels are split by the $T_d$ crystal field into  into $e$ and $t_2$ levels. Mn $t_2$ levels hybridize with the NN carbon $2p$ levels to form bonding $t_B$ and antibonding $t_{AB}$ hybrid levels. The bonding $t_{B}$ levels are nearly spin degenerate and form a crystal field resonance state located around -2.2 eV below the VBM. The spin up and spin down $e$ levels are localized on the Mn site, giving rise to two bound states situated at 0.6 eV and 1.8 eV above the VBM  separated by a Hund exchange splitting $J^H=1$ eV. The Mn $e$ level contribution to the total magnetization is about 70 \% and the bonding $t_B$ states are weakly spin-polarized. The antibonding $t_{AB}$ levels are weakly spin-polarized and form a bound state situated at 5.1 eV above the VBM. The number of electrons located on the Mn ion is 23.5, which corresponds to a $4s^03d^5$ configuration and spin $S_1=\frac{1}{2}$. Mn$^{2+}$ induces two extra electrons distributed over the NN carbon dangling bonds with  $2s^22p^4$ configuration and spin $S_2=0$. The Mn $(t_2)^2$ level hybridizes with the  NN carbon $2p^4$ $t_2$ defect level. The bonding $t_B$ states are fully occupied and, as in the case of Mn, the $e$ levels are partially occupied in the configuration $(e^\uparrow)^2(e^\downarrow)^1$, corresponding to a total magnetic moment $M_T=1.0\ \mu_\mathrm{B}$ entirely localized on the Mn. Fig. \ref{Mn} explains the observed DOS by a $p$--$d$ hybridization model.

\begin{figure}[h!]
\begin{tabular}{|c|}
\hline
\includegraphics[width = 85mm]{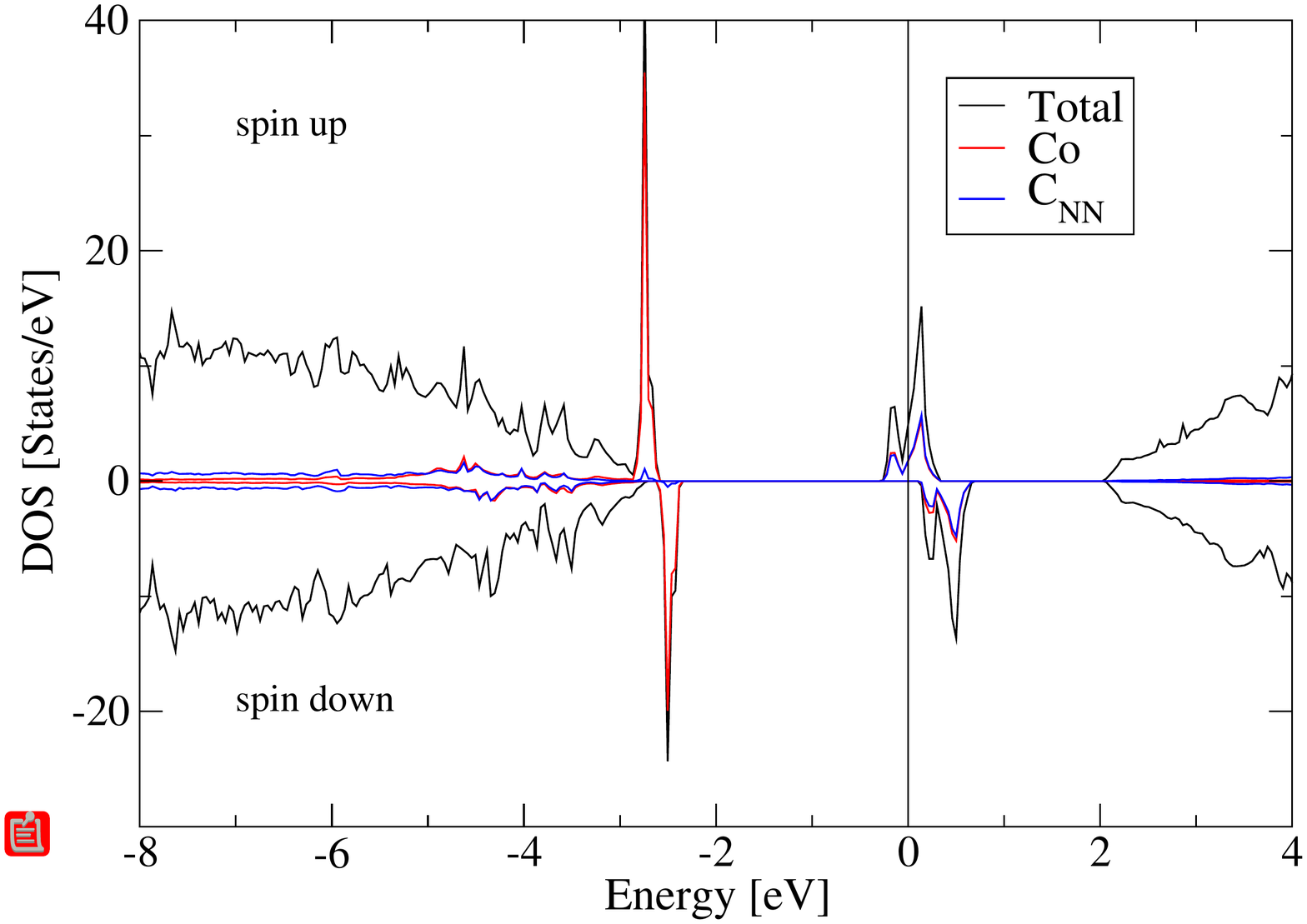}\\\hline  
 \\
\includegraphics[width = 80mm]{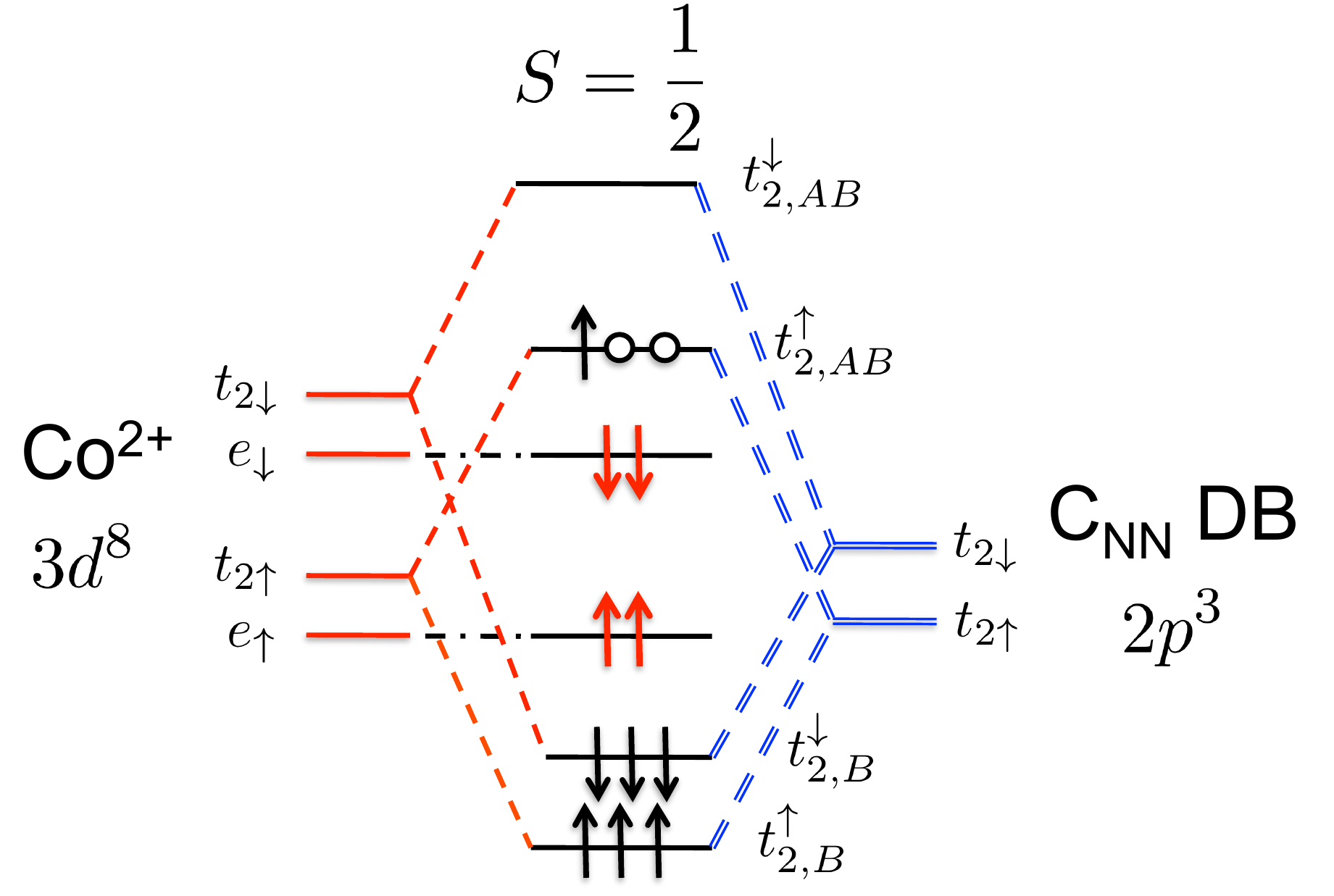}\\  
 \\\hline
 
 \end{tabular}
\caption{(color online) Top : Total, Co and NN C partial DOS. The antibonding state $t_{AB}$ has approximately a 50 \% Co $3d$ character (50 \% NN C $2p$ character). Bottom : Schematic representation of the $p$--$d$ hybridization model for Co$^0_s$. Symbols are the same as in Fig.~\ref{Ti}.}\label{Co} 
\end{figure}

Fig. \ref{Co} shows the total, Co and NN carbon partial DOS of the CoC$_{63}$ supercell. The Co $3d$ levels are split by the $T_d$ crystal field  into $e$ and $t_2$ levels. The Co $t_2$ levels hybridize with the NN carbon $2p$ levels to form bonding $t_B$ and antibonding $t_{AB}$ hybrid levels. The bonding $t_{B}$ levels are weakly spin-polarized and form a crystal field resonance state located around -2.7 eV below the VBM. The spin up and spin down $e$ levels are localized on the Co site, giving rise to two peaks at the top of the valence band. The number of electrons located on the Co ion is 25.9 which corresponds to a Co$^+$ configuration $4s^03d^8$ and a spin $S_1=0$. Co$^{+}$ induces one extra electron distributed over NN carbon dangling bonds of configuration $2s^22p^3$ and spin $S_2=\frac{1}{2}$. The Co $(t_2)^4$ level hybridizes with the  NN carbon $2p^3$ $t_2$ defect level. The $2p^3$ configuration of the NN carbon dangling bond is responsible for the $S=\frac{1}{2}$ spin ($M_T=1.0\ \mu_\mathrm{B}$) which is distributed over the Co ($M_\mathrm{Co}=0.6\ \mu_\mathrm{B}$) and the 4 NN carbon atoms ($M_\mathrm{C}=0.1\ \mu_\mathrm{B}$). Fig. \ref{Co} explains the observed DOS by a $p$--$d$ hybridization model.

\subsection{The $S=1$ centers : Cr$^0_s$ and Ni$^0_s$.}

\begin{figure}[h!]
\begin{tabular}{|c|}
\hline
\includegraphics[width = 85mm]{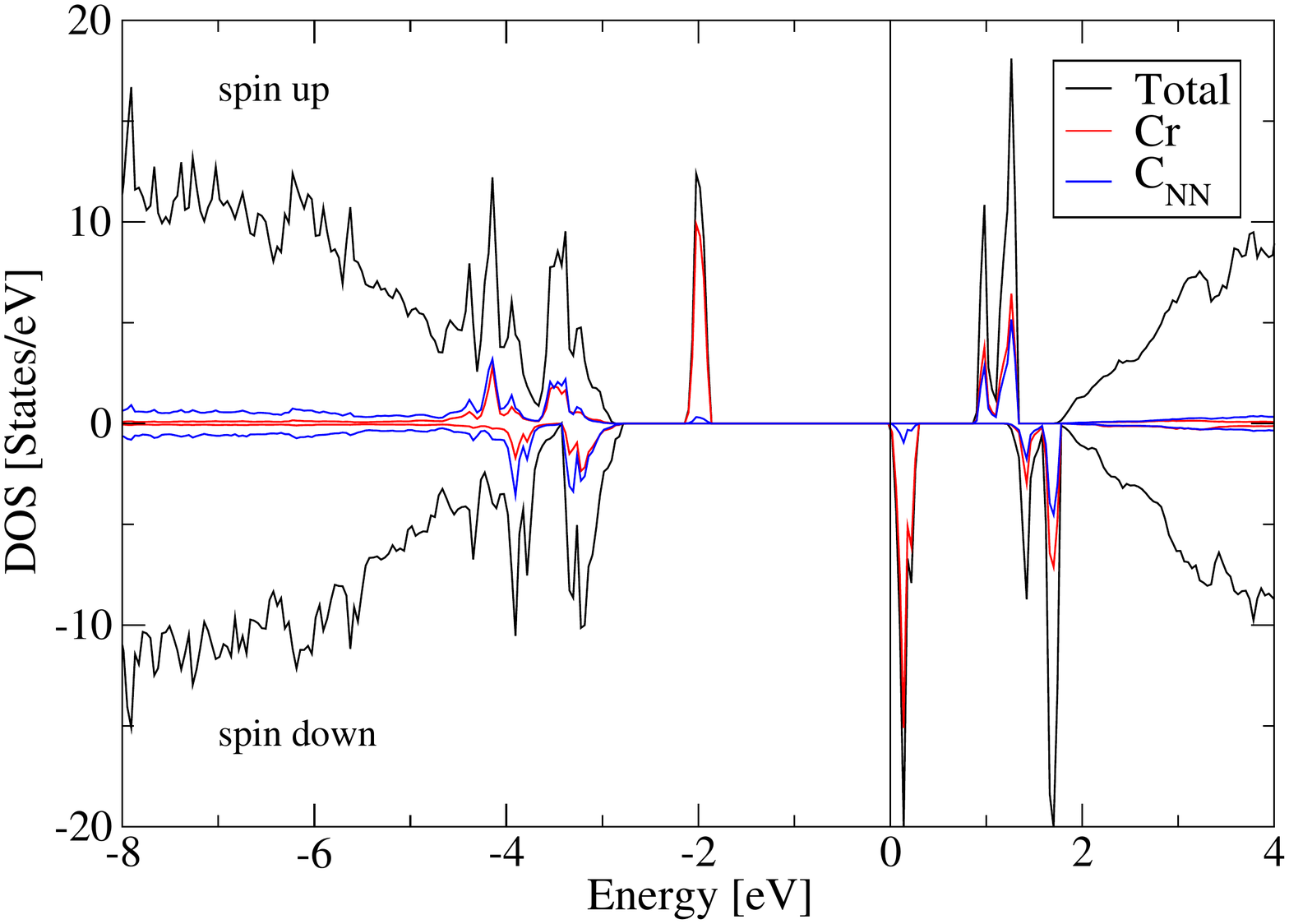}\\\hline  
 \\
\includegraphics[width = 80mm]{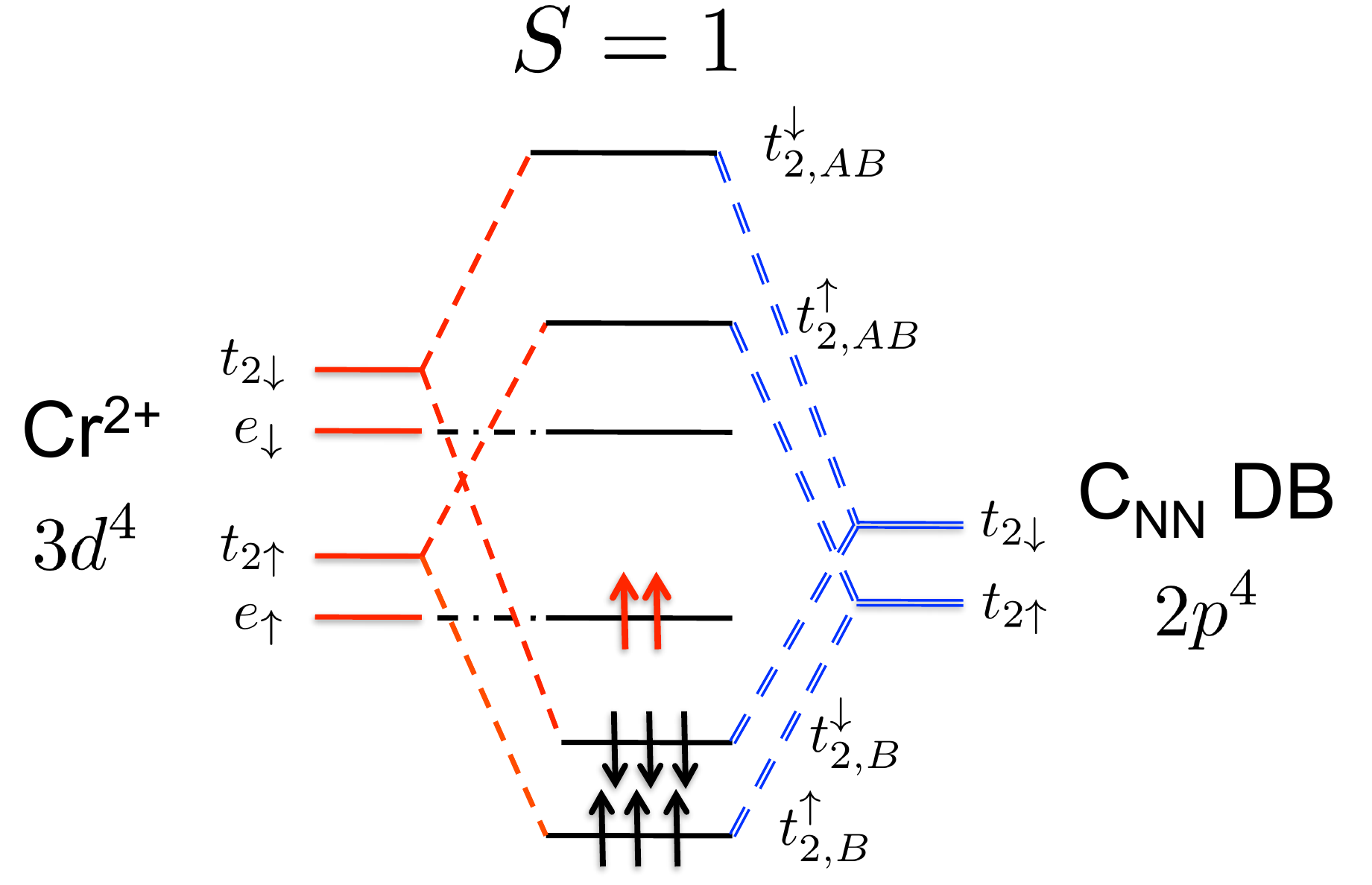}\\  
 \\\hline
\end{tabular}
\caption{(color online) Top : Total, Cr and NN C partial DOS. Bottom : Schematic representation of the $p$--$d$ hybridization model for Cr$^0_s$. Symbols are the same as in Fig.~\ref{Ti}.}\label{Cr} 
\end{figure}

Fig. \ref{Cr} shows the total, Cr and NN carbon partial SGGA DOS of the CrC$_{63}$ supercell. The Cr $3d$ levels are split by the $T_d$ crystal field  into $e$ and $t_2$ levels. Cr $t_2$ levels hybridize with the NN carbon $2p$ levels to form bonding $t_B$ and antibonding $t_{AB}$ hybrid levels. The bonding $t_{B}$ levels are nearly spin-degenerate and form crystal field resonance states around -1.5 eV below the VBM. The spin up and spin down $e$ levels are localized on the Cr site, giving rise to two bound states approximately  0.8 eV and 2.8 eV above the VBM and separated by a Hund exchange splitting $J^H=2$ eV. The Cr $e$ level contribution to the total magnetization is roughly 80 \% of the total magnetization and the bonding $t_B$ states are weakly spin-polarized.  The antibonding $t_{AB}$ levels are totally empty and forms two bound states around 4.3 eV above the VBM. The number of electrons located on the Cr ion is 22.3, which correspond to a Cr$^{2+}$ configuration $4s^03d^4$ and spin $S_1=1$. Cr$^{2+}$ induces two extra electrons distributed over NN carbon dangling bonds of configuration $2s^22p^4$ and spin $S_2=0$. The Cr $(t_2)^2$ level hybridizes with the  NN carbon $2p^4$ $t_2$ defect level. The bonding $t_B$ states are fully occupied and  the $e$ levels are partially occupied in the configuration $(e^\uparrow)^2(e^\downarrow)^0$ corresponding to a total magnetic moment $M_T=2.0\ \mu_\mathrm{B}$ entirely localized on the Cr ion. Fig. \ref{Cr} explains the observed DOS by a $p$--$d$ hybridization model.

\begin{figure}[h!]
\begin{tabular}{|c|}
\hline
\includegraphics[width = 85mm]{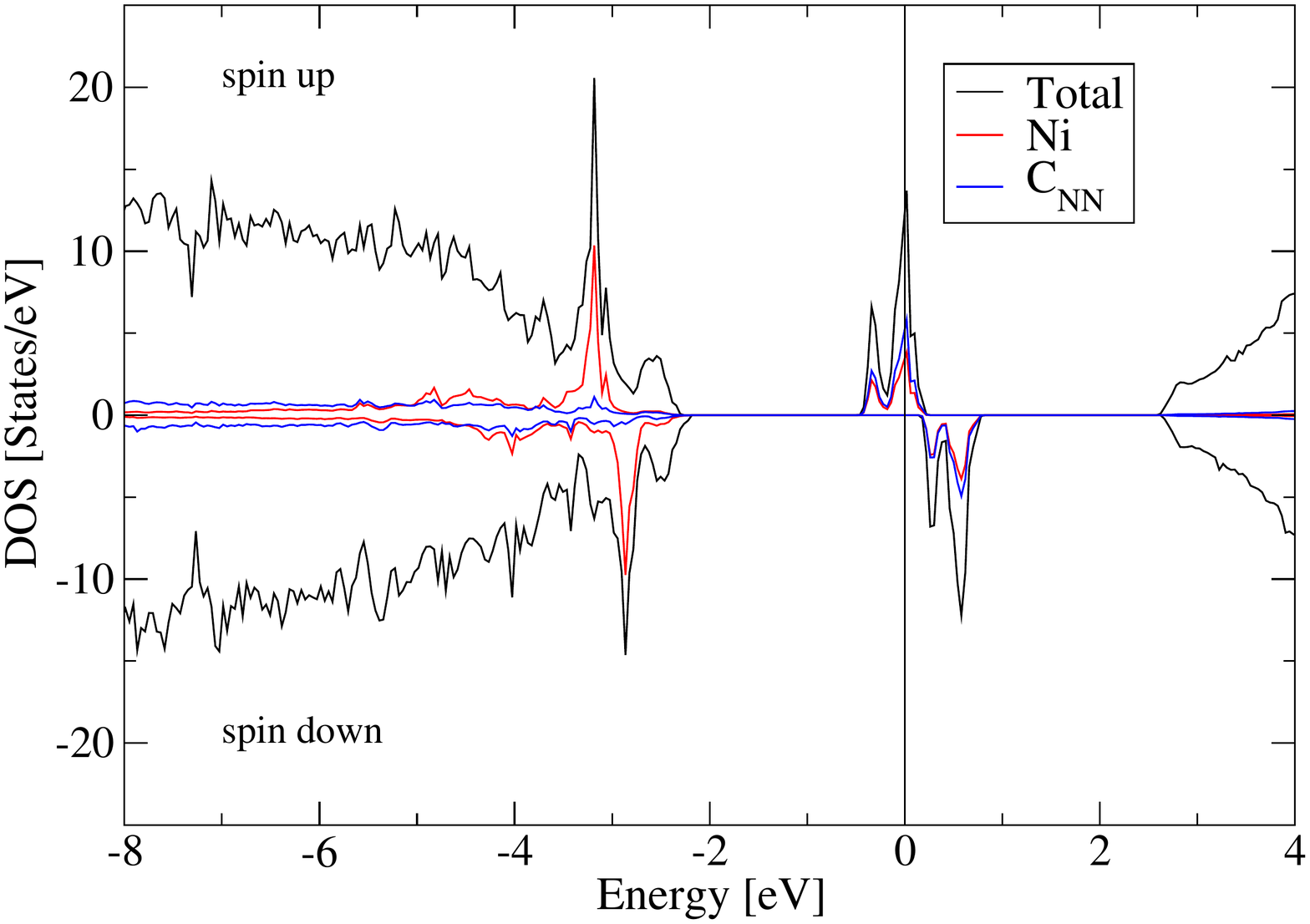}\\\hline  
 \\
\includegraphics[width = 80mm]{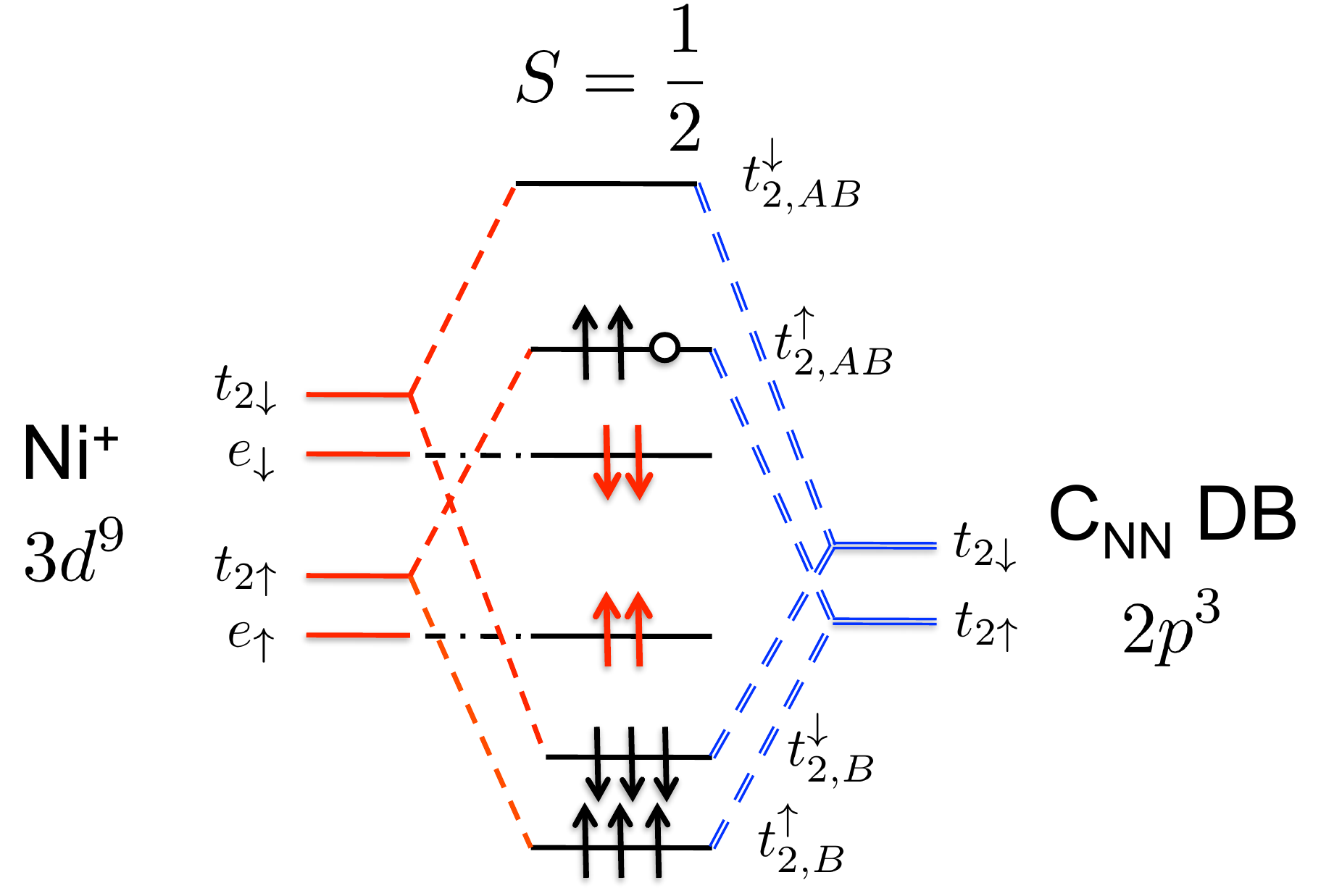}\\  
 \\\hline
 
 \end{tabular}
\caption{(color online) Top : Total, Ni and NN C partial DOS.  Bottom : Schematic representation of the $p$--$d$ hybridization model for Ni$^0_s$. Symbols are the same as in Fig.~\ref{Ti}.}\label{Ni} 
\end{figure}

Fig. \ref{Ni} shows the total, Ni and NN carbon partial SGGA density of states (DOS) of the NiC$_{63}$ supercell. The Ni $3d$ levels are split by the $T_d$ crystal field  into $e$ and $t_2$ levels. Ni $t_2$ levels hybridize with the NN carbon $2p$ levels to form bonding $t_B$ and antibonding $t_{AB}$ hybrid levels.The spin up and spin down bonding $t_{B}$ levels forms a crystal field resonance state situated around -2.1 eV below the VBM. The spin up and spin down $e$ levels are localized on the Ni site, giving rise to two peaks approximately 1 eV below the valence band maximum (VBM) and  separated by a Hund exchange splitting $J^H=0.4$ eV. The spin up and spin down antibonding $t_{AB}$ levels are located at 2.5 and 3.0 eV above the VBM. The antibonding $t_{AB}$ level contribution to the total magnetization is more than 70 \%, with 50 \% from Ni $3d$ level and 20 \% from the NN carbon $2p$ level. The number of electrons located on the Ni ion is 26.9, which corresponds to a Ni$^+$ configuration $4s^03d^9$ with a spin $S_1=\frac{1}{2}$. Ni$^{+}$ induces one extra electron distributed over the four NN carbon dangling bonds of configuration $2s^22p^3$ and spin $S_2=\frac{1}{2}$. The Ni $t_2$ levels hybridize with the  NN carbon $2p^3$ $t_2$ defect level. This $p$--$d$ hybridization picture identifies the origin of the calculated SGGA DOS as due to a FM interaction between $S_1$ and $S_2$ as shown on Fig. \ref{Ni}.

\subsection{The unusual case of Cu$^0_s$ : a $S=\frac{3}{2}$ center.}

\begin{figure}[h!]
\begin{tabular}{|c|}
\hline
\includegraphics[width = 85mm]{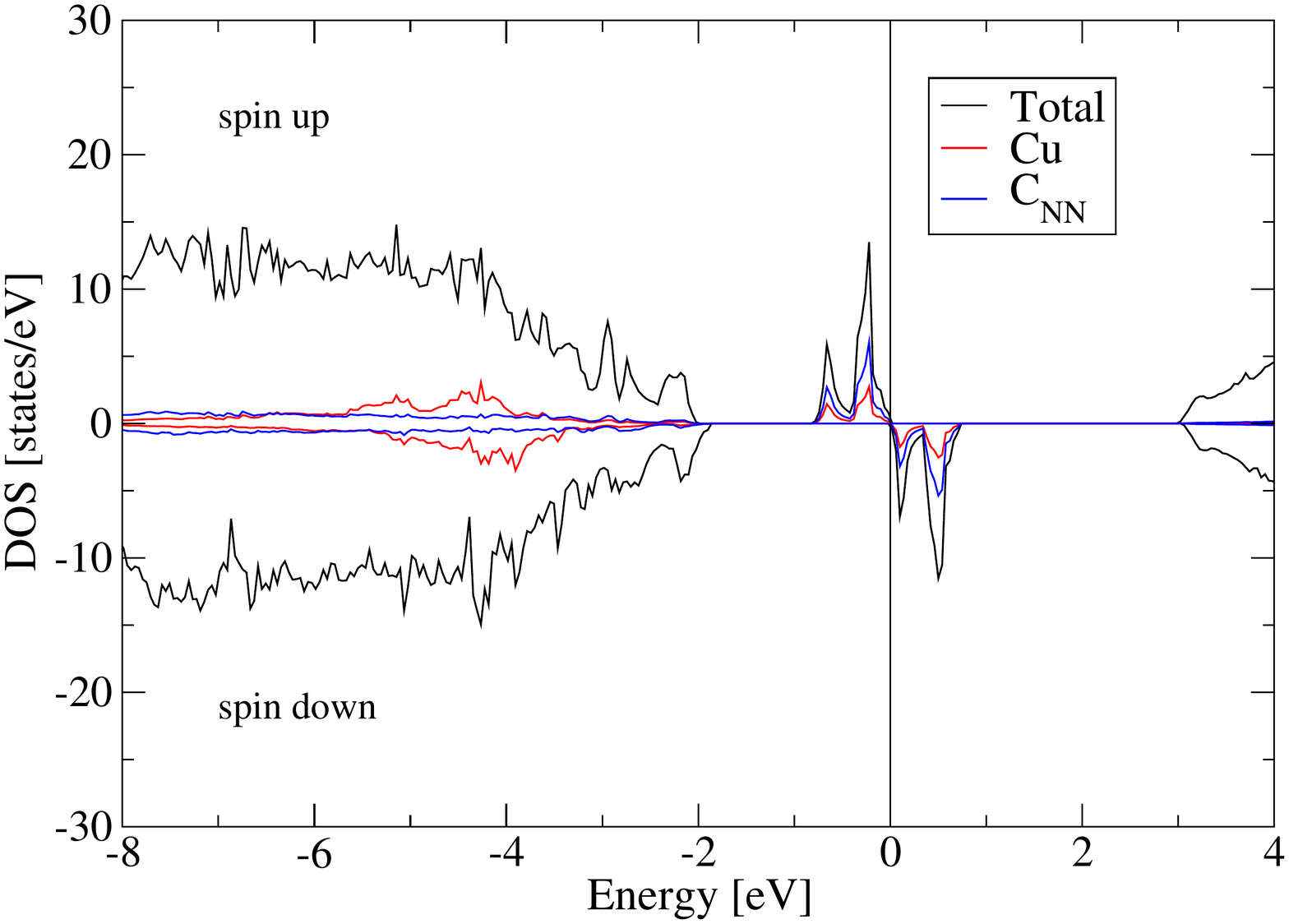}\\\hline  
 \\ 
\includegraphics[width = 80mm]{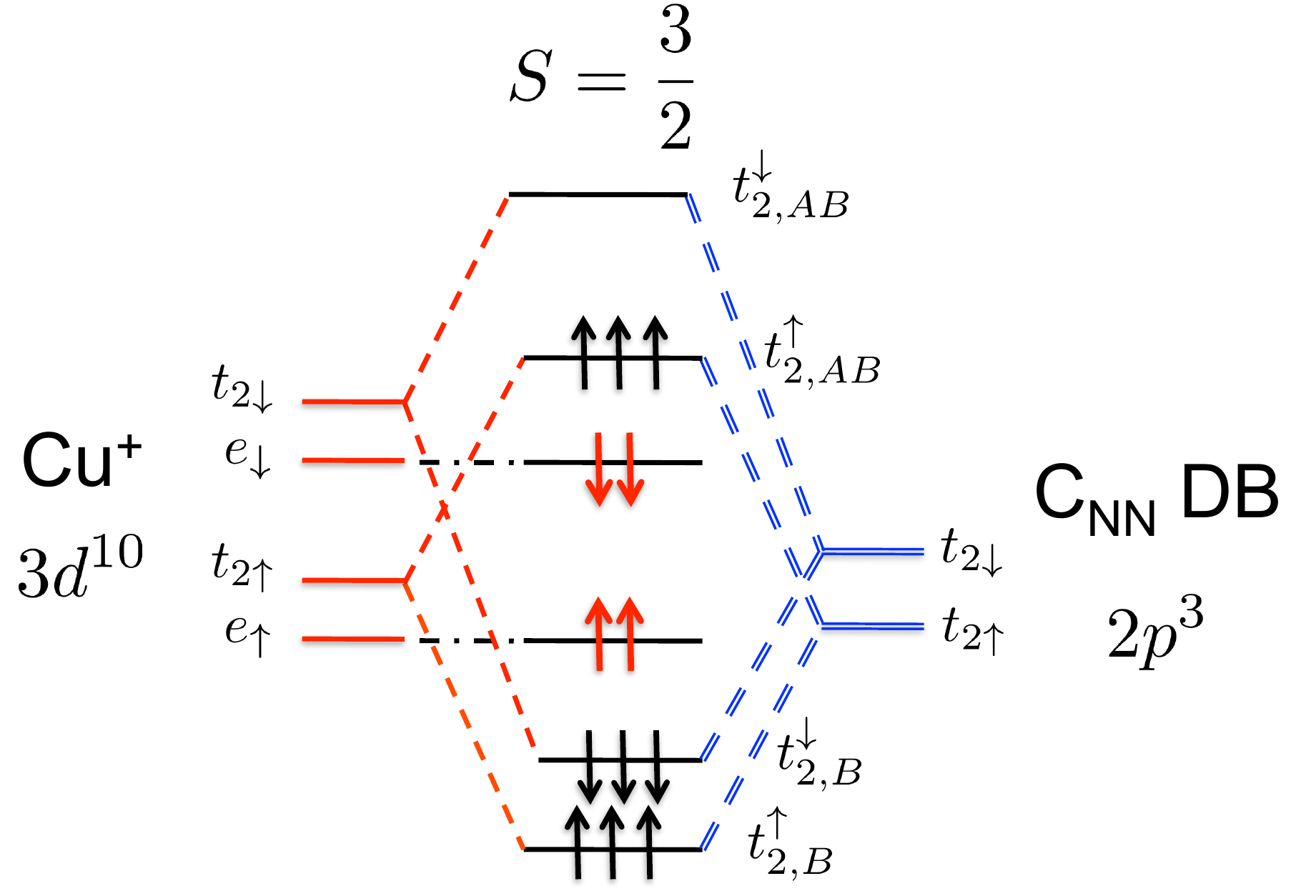}\\  
 \\\hline
 
 \end{tabular}
\caption{(color online) Top : Total, Cu and NN C partial DOS. Bottom : Schematic representation of the $p$--$d$ hybridization model for Cu$^0_s$. Symbols are the same as in Fig.~\ref{Ti}.}\label{Cu} 
\end{figure}

Fig. \ref{Cu} shows the total, Cu and NN carbon partial DOS of the CuC$_{63}$ supercell. The Cu $3d$ levels are split by the $T_d$ crystal field  into $e$ and $t_2$ levels. Cu $t_2$ levels hybridize with the NN carbon $2p$ levels to form bonding $t_B$ and antibonding $t_{AB}$ hybrid levels. The bonding $t_{B}$ levels are weakly spin-polarized and are located around 2.6 eV below the VBM. The spin up and spin down $e$ levels are localized on the Cu site, giving rise to two nearly spin-degenerate peaks situated around 2.4 eV below the VBM. The spin up and spin down antibonding $t_{AB}$ levels are situated at 1.9 and 2.7 eV above the VBM. The number of electrons located on the Cu ion is 28.1, which corresponds to a Cu$^+$ configuration $4s^03d^{10}$ and a spin $S_1=0$. Cu$^{+}$ induces one extra electron distributed over the NN carbon dangling bonds of configuration $2s^22p^3$ and spin $S_2=\frac{3}{2}$. The Cu $t_2$ level hybridizes with the NN carbon $2p^3$ $t_2$ defect level. The $2p^3$ configuration of the NN carbon dangling bond is responsible for the total spin $S=\frac{3}{2}$ ($M_T=3.0\ \mu_\mathrm{B}$), which is distributed over the Cu ($M_\mathrm{Cu}=0.8\ \mu_\mathrm{B}$) and the 4 NN carbon atoms ($M_\mathrm{C}=0.4\ \mu_\mathrm{B}$). Fig. \ref{Cu} explains the observed DOS by a $p$--$d$ hybridization model. Cu$^+$ $3d$ levels are nearly spin degenerate corresponding to  a non-magnetic $3d^{10}$ closed-shell configuration.

\subsection{Outliers: Sc$^0_s$ ($S=\frac{1}{2}$) and Zn$^0_s$ ($S=1$).}

\begin{figure}[h!]

\includegraphics[width = 85mm]{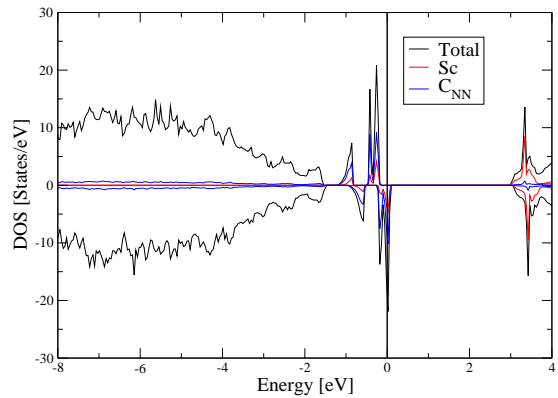}
\caption{(color online) Total, Sc and NN C partial DOS.}\label{Sc} 
\end{figure}

Fig. \ref{Sc} shows the total, Sc and NN carbon partial SGGA density of states (DOS) of the ScC$_{63}$ supercell. Sc $3d$ levels hybridize slightly with the NN carbon $t_2$ levels and give rise to 2 spin-up and 2 spin-down bound states situated at 0.7 and 1.2 eV (1.0 and 1.6 eV). The spin-down $t_2$ level at 1.6 eV is partially empty (1 hole) which is responsible for the total spin-polarization $M_T=1\ \mu_\mathrm{B}$ which is distributed on the Sc and NN carbon atoms. The two spin-degenerated peaks at 4.6 eV arises from the Sc $e$ levels. There are no antibonding levels and the density of states cannot be explained by the $p-d$ hybridization model used above for Ti through Cu.

\begin{figure}[h!]
\includegraphics[width = 85mm]{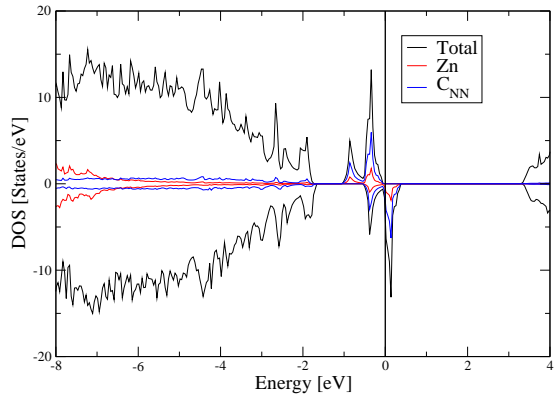}
\caption{(color online) Total, Zn and NN C partial DOS.}\label{Zn} 
\end{figure}

Fig. \ref{Zn} shows the total, Zn and NN carbon partial SGGA density of states (DOS) of the ZnC$_{63}$ supercell. Zn gives rise to two bound states of $t_2$ symmetry at 1.6 and 2.1 eV above the VBM which are almost entirely of NN carbon $2p$ character. The partially empty (2 holes) bound state at 2.1 eV above the VBM is responsible for the total magnetization $M_T=2\ \mu_\mathrm{B}$ which is distributed over the Zn and NN carbon atoms. The number of electron on Zn is 29.0 corresponding to a configuration $3d^{10}4s^1$. Thus the Zn $4s$ level is almost depleted and the Zn $4s$ electron occupies a bound state of mostly NN carbon $2p$ character to lower its energy. The $Zn^+$ $3d$ levels are delocalized and are located at about -6 eV below the top of the valence band and are nearly spin degenerate, corresponding to  a non-magnetic $3d^{10}$ closed-shell configuration.

\section{High-Spin to Low-Spin Transition with Strain}\label{highspin-lowspin}

\begin{figure}[h!]
\includegraphics[width = 85mm]{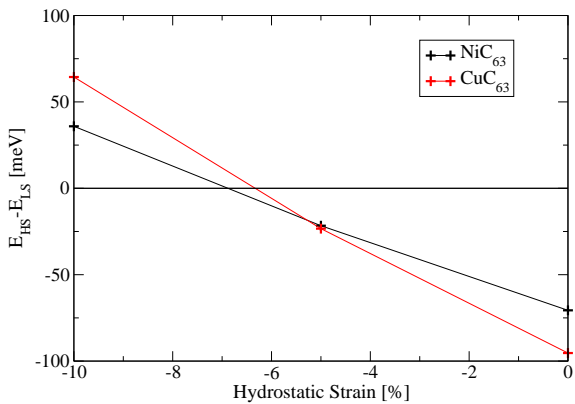}
\caption{Total energy difference between low-spin and high-spin fixed spin moment SGGA calculations as a function of hydrostatic strain.}\label{HSvsLS} 
\end{figure}

We now study the effect of hydrostatic pressure on the magnetic properties of high-spin TM$^0_s$ dopants (TM = Cr, Ni and Cu). For that purpose, we take the total energy difference between high-spin (HS) and low-spin (LS) fixed-spin calculations. The HS (LS) state of Cr$^0_s$ and Ni$^0_s$ corresponds to a total magnetic moment $M_\mathrm{HS}=2\ \mu_\mathrm{B}$ ($M_\mathrm{LS}=0\ \mu_\mathrm{B}$) whereas the HS (LS) state of Cu$^0_s$ corresponds to $M_\mathrm{HS}=3\ \mu_\mathrm{B}$ ($M_\mathrm{LS}=1\ \mu_\mathrm{B}$). For Cr$^0_s$, we do not obtain a transition due to the large magnetic energy of the $S=1$ ground state, about 1 eV energetically lower than the $S=0$ non-magnetic solution. The energetic stability of the magnetic ground state is associated with the $e$ character of the highly spin-polarized states, whereas both Ni and Cu have magnetism driven by the $t_2$ states. The results for Ni$^0_s$ and Cu$^0_s$ are presented in Fig. \ref{HSvsLS}. We obtain a transition from high-spin $S=1$ ($S=\frac{3}{2}$) to low-spin  $S=0$ ($S=\frac{1}{2}$) under compressive hydrostatic strain with a transition at $e_H=-7$ \% ( $e_H=-6$ \%) for Ni$^0_s$ (Cu$^0_s$).

\begin{figure}[h!]
\begin{tabular}{|c|}
\hline
\includegraphics[width = 85mm]{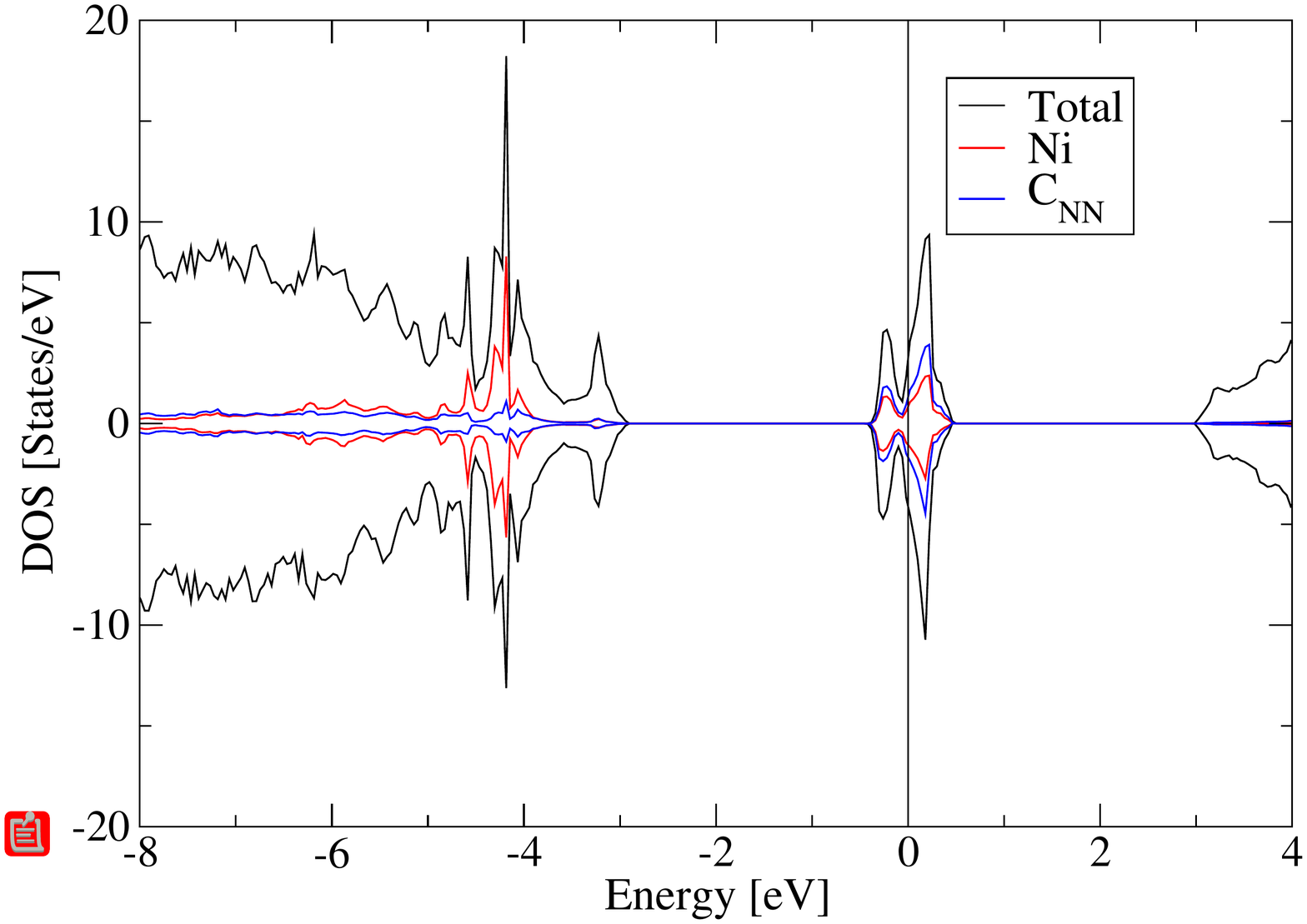}\\\hline  
 \\
\includegraphics[width = 80mm]{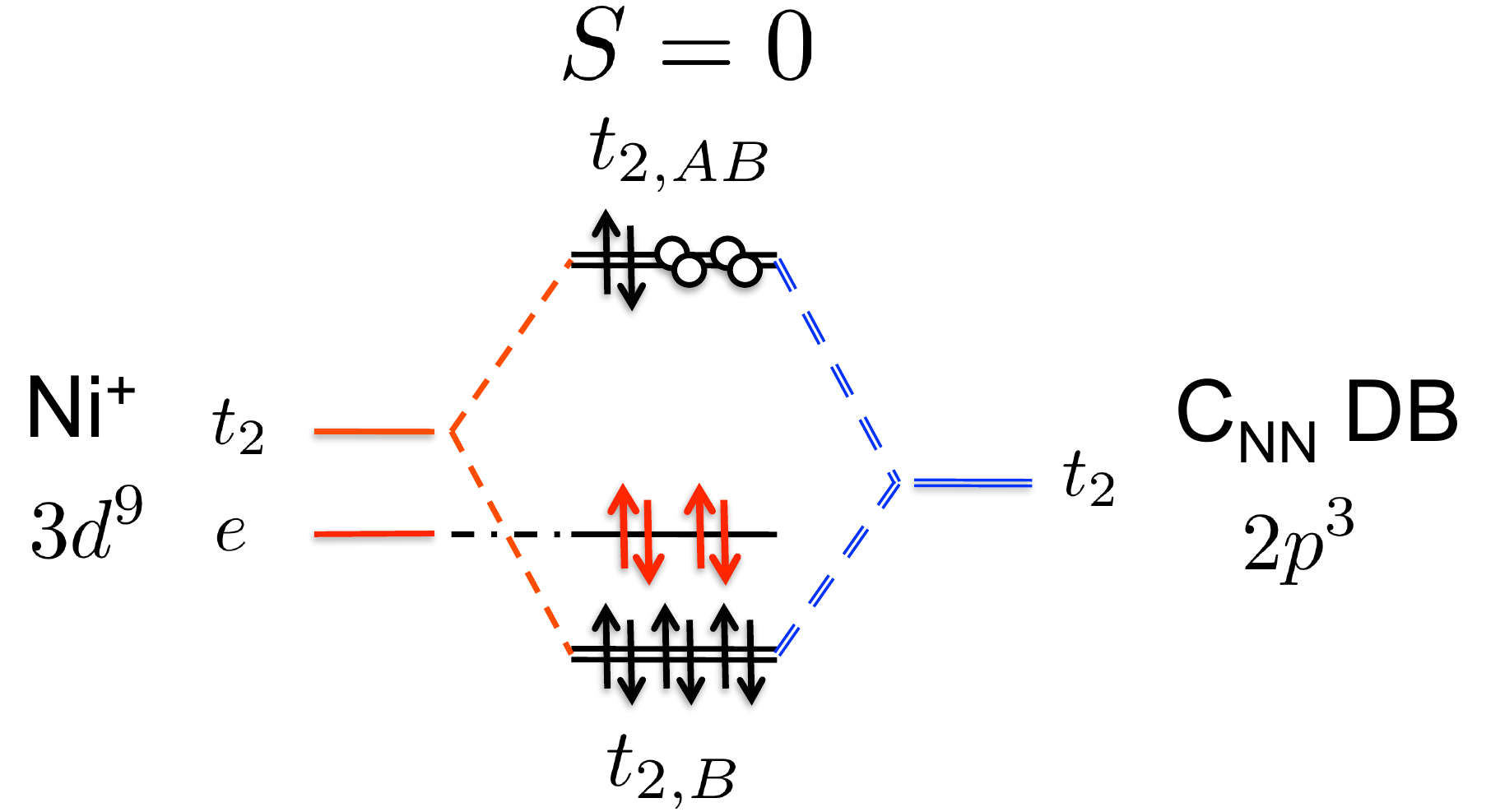}\\  
 \\\hline
 
 \end{tabular}
\caption{Low-spin state of Ni$^0_s$, $S=0$ : Top : Total and Ni and NN C partial DOS. Bottom : Schematic representation of the $p-d$ hybridization model for Ni$^0_s$.}\label{NiLS} 
\end{figure}

\begin{figure}[h!]
\begin{tabular}{|c|}
\hline
\includegraphics[width = 85mm]{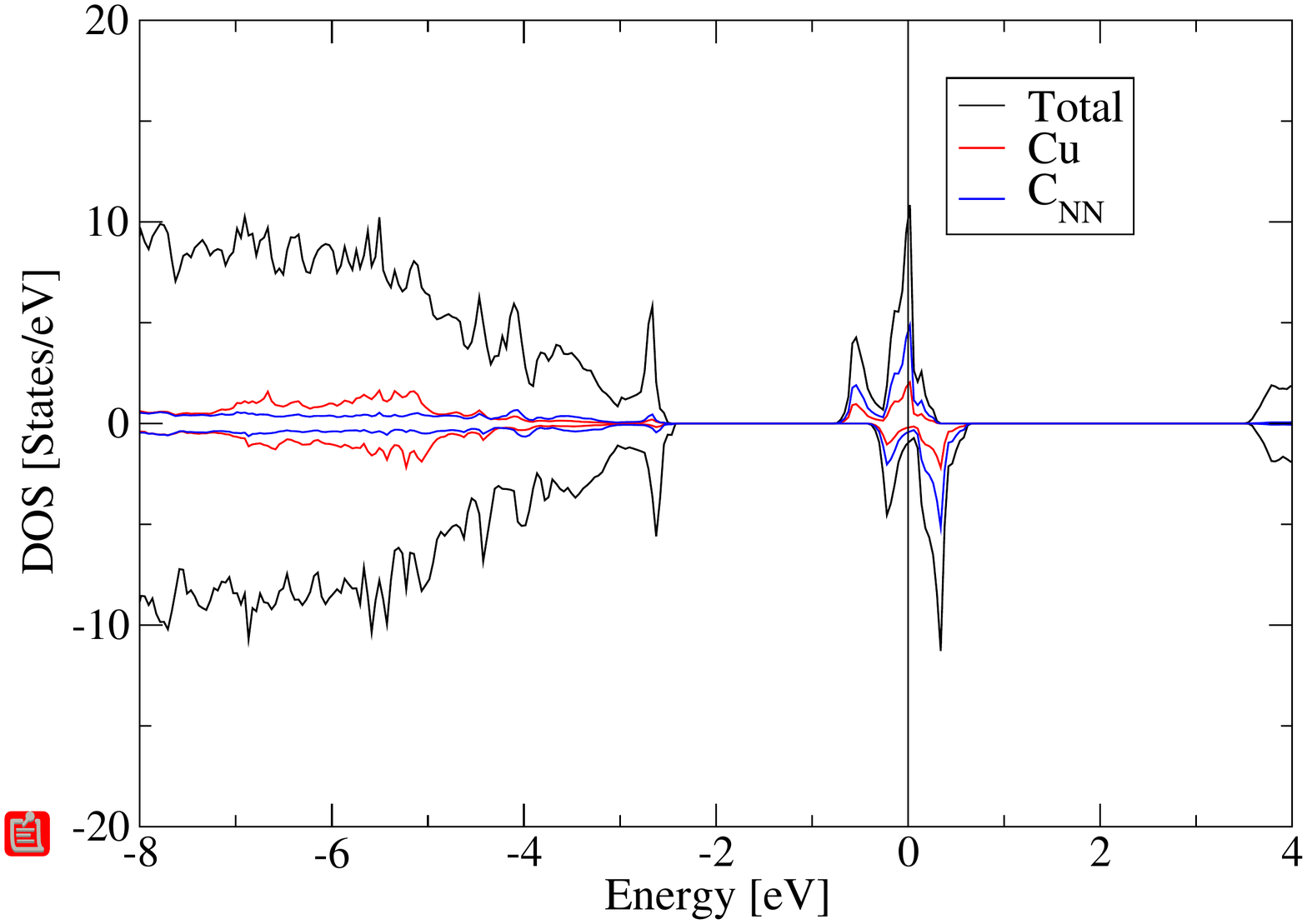}\\\hline  
 \\
\includegraphics[width = 80mm]{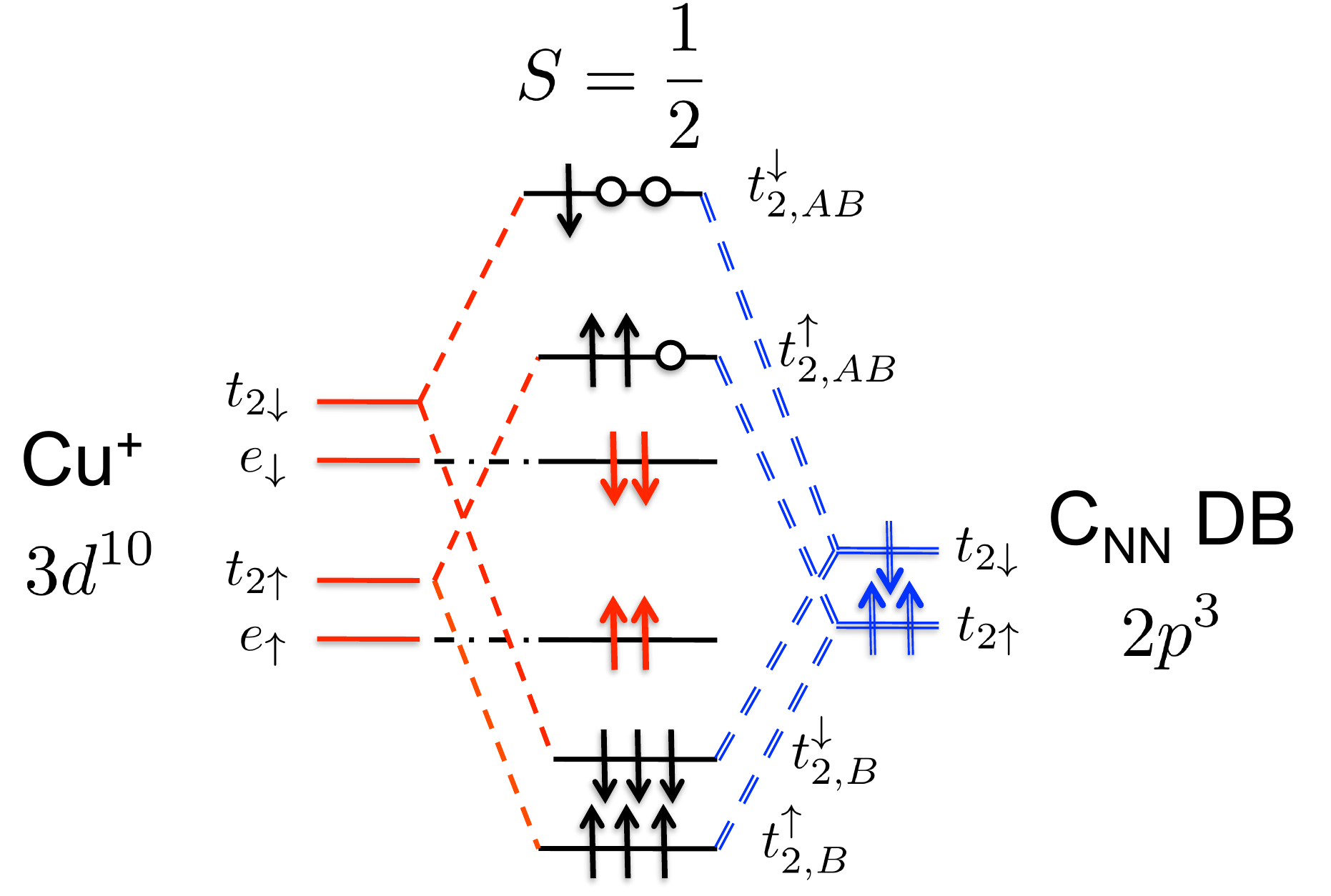}\\  
 \\\hline
 
 \end{tabular}
\caption{Low-spin state of Cu$^0_s$, $S=\frac{1}{2}$ : Top : Total and Cu and NN C partial DOS. Bottom : Schematic representation of the $p-d$ hybridization model for Cu$^0_s$.}\label{CuLS} 
\end{figure}

Fig. \ref{NiLS} and \ref{CuLS} presents the DOS of the low-spin SGGA ground state of Ni$^0_s$ and Cu$^0_s$ under compressive hydrostatic pressure $e_H=-10$ \%. The corresponding total magnetic moment for the (spin unfixed) SGGA solution is $M_\mathrm{LS}=0.1\ \mu_\mathrm{B}$ for Ni$^0_s$  and $M_\mathrm{LS}=1.4\ \mu_\mathrm{B}$ for Cu$^0_s$. The TM$^+$ configuration $4s^03d^{n+1}$ remains unchanged, which induces an extra electron on the NN carbon atoms, creating a $2s^22p^3$ defect level. Fig. \ref{NiLS} and \ref{CuLS} explain the observed DOS with a $p-d$ hybridization model. For Ni$^0_s$, the spin-degenerate Ni $3d$ levels hybridize with the spin-degenerate $2p^3$ dangling bonds giving rise to a non-magnetic interaction, corresponding to an antiferromagnetic interaction between the Ni $S_1=\frac{1}{2}$ and a spin $S_2=-\frac{1}{2}$ distributed over the NN carbon dangling bonds.  For Cu$^0_s$, the spin-degenerate Cu $3d$ levels hybridize with the $2p^3$ dangling bond. The $2p^3$ configuration is responsible for the total spin $S=\frac{1}{2}$ distributed over the NN carbon dangling bonds. 

The nearly-degenerate $S=1$ and $S=0$ states of the substitutional nickel impurity in diamond can be understood as an exchange-coupled system of two electron spins: one localized on the nickel ion and one delocalized on the four nearest-neighbor (NN) carbon atoms\cite{Chanier2011a}. Strain changes the exchange coupling and as a result modifies the ground state from parallel to antiparallel alignment. A similar phenomenon occurs for Cu, but the transition is between $S=3/2$ and $S=1/2$.
The existence of a pressure induced transition between a high-spin and low-spin state is also observed in transition-metal-ion compounds, where it was shown theoretically to be due to a competition between localization (induced by the Hund's exchange coupling which favors the high-spin state) and a tendency towards delocalization induced by the crystal field (which favors the low-spin state)\cite{Bari1972}.

\section{Conclusion}\label{conclude}

We have studied by first-principle calculations the chemical trends of  neutral substitutional transition-metal dopants in diamond. The origin of the magnetism for these dopants has been explained by a $p$--$d$ hybridization model. Ti and Fe impurities are non-magnetic corresponding to closed bonding hybridized $t_2$ and an empty (Ti) or full (Fe)  $e$ level respectively. For V, Cr and Mn, the magnetization is entirely localized on the TM ion and driven by the $e$ levels. For Co, Ni and Cu, the magnetization is distributed over the TM ion and the NN carbon atoms with a magnetization due to the antibonding $t_2$ levels. Electron paramagnetic study of these magnetic centers would be of great interest in order to confirm the calculated chemical trend.

The calculated total magnetic moments for the TM ion series are in agreement with previous GGA calculations\cite{Assali2009}. Ref.~\onlinecite{Assali2009} finds the same relaxation with point goup $T_d$ symmetry for Ti, Cr and Fe. They found a NN relaxation of $D_{2d}$ for V, Mn and Co and a $C_1$ symmetry relaxation for Ni. To verify the validity of our results,  which were all-atom relaxation calculations performed with $T_d$ symmetry, we did nearest-neighbor atom relaxation  calculations  within GGA (1 meV/\AA\ precision) with no symmetry constraints, fixing the other atomic positions to their ideal value. For Ni, the tetrahedra calculated this way is slightly distorted to $C_1$ symmetry, however the distortion is less than a 1\% difference in distance and less than a 0.1\% difference in angle, with a magnetic energy $E_M=48$ meV and a total spin $S=1$. Hence our $T_d$-symmetry results represent well the state of these transition-metal ions.

The nearly-degenerate $S=1$ and $S=0$ states of the substitutional nickel impurity in diamond can be understood as an exchange-coupled system of two electron spins: one localized on the nickel ion and one delocalized on the four nearest-neighbor (NN) carbon atoms. Such nearly-degenerate $S=1$ and $S=0$ states can be used to construct an effective two-state decoherence free subspace from the  $S_z=0$ triplet  ($T_0$) and the singlet ($S$) states. This approach would be similar to that implemented for double quantum dots \cite{Petta2005} with electrostatic gating. For the nickel ion, strain modulation could be used to manipulate the energy splitting between the $S$ and the $T_0$ states,  instead of the electrostatic gating used for double quantum dots \cite{Petta2005}. This is one way that the electronic configurations of transition-metal dopants could be used to encode quantum information in a form amenable for solid state quantum information processing.

This work was supported by DARPA QuEST.

\bibliography{central-bibliography}




\end{document}